\documentclass[
 preprint, 
 amsmath,amssymb,
 aps, physrev,
]{revtex4-2}
\usepackage{graphicx}
\usepackage{amssymb}
\usepackage{amsmath}
\usepackage{comment}
\usepackage{subcaption}
\usepackage{xcolor}
\usepackage[group-separator={,}]{siunitx}
\usepackage{bm}

\begin{document}

\preprint{APS/123-QED}

\title{Generation of an isolated vortex gust through a heaving and pitching foil}

\author{Bingfei Yan}\thanks{These authors contributed equally to this work.}
\affiliation{%
Department of Mechanical Engineering, University of Wisconsin--Madison, Madison, WI
}

\author{Eric Handy-Cardenas}\thanks{These authors contributed equally to this work.}
\affiliation{%
 School of Engineering, Brown University, RI
}%

\author{Kenny Breuer}
\affiliation{%
 School of Engineering, Brown University, RI
}%

\author{Jennifer A. Franck}
\affiliation{%
Department of Mechanical Engineering, University of Wisconsin--Madison, Madison, WI, US
}
\email{Contact author: jafranck@wisc.edu}

\date{\today}

\begin{abstract}
This study introduces a vortex gust generation method for isolated vortices impacting a downstream airfoil that is applicable to both numerical simulations and experiments. The vortex gust is generated by a symmetric airfoil undergoing a rapid pitching maneuver during a prescribed heaving motion. The resulting vortices propagate along trajectories nearly parallel to the incoming flow, while the associated wake extends obliquely from the vortex core. Despite differences in Reynolds number, rapid pitching duration and detailed vortex structure between simulations and experiments, consistent trends are observed in how the vortex rotation orientation, strength, and position vary with the prescribed motion parameters. Analysis of the lift response of the downstream airfoil shows that the aerodynamic influence associated with the wake does not persist over extended time scales. These results demonstrate that the proposed method enables the controlled generation of vortex gusts with prescribed characteristics, providing a flexible approach for systematic studies of vortex–airfoil interaction.
\end{abstract}

\maketitle
\section{Introduction}
\label{sec:intro}

Gusts are spatially and temporally varying flow disturbances that are commonly encountered in both natural and engineered environments. In applications ranging from unmanned aerial vehicles (UAVs) to wind and tidal turbines, lifting surfaces are commonly exposed to complex atmospheric or wake-driven gust environments, where the characteristic time and length scales of unsteadiness may be comparable to those governing the aerodynamic response of the structure. These unsteady flow features can induce large transient aerodynamic loads, posing significant challenges to the performance, stability, and structural integrity of flight vehicles and energy-harvesting systems \cite{jones2022physics,mohamed2023gusts}. Canonical gust models have been widely employed to study these interactions in a controlled and systematic manner. This research focuses on the model of a vortex gust interacting with a airfoil. Specifically, this manuscript introduces a method to generate predictable and isolated vortex gusts across both experiments and simulation, that are not impacted by the surrounding wake structure of the vortex generator. 

Classical aerodynamic theory has primarily focused on transverse and streamwise gust models \cite{kussner1932stresses,von1938airfoil,miles1956aerodynamic,atassi1984sears}, which can be conveniently analyzed using linearized potential flow formulations. In contrast, vortex gusts, characterized by compact and coherent regions of concentrated vorticity, are more difficult to describe within such frameworks because they are highly localized, directional, and inherently transient. As Jones et al. \cite{jones2020gust} notes, a central challenge in gust response research lies in isolating a well-defined interaction within an inherently unsteady, multiscale flow. In this context, the method by which a vortex gust is generated becomes crucial, as it determines both the structure of the incoming disturbance and the degree to which the resulting interaction can be systematically analyzed. This study and the following review of previous work focus specifically on vortex gusts consisting of a single dominant and isolated vortex.
\par
Current studies of vortex-body interactions utilize two primary methods for generating isolated vortex gust. The first type, only possible in numerical studies, involves directly imposing a theoretically modeled traveling vortex upstream of the body that interacts with the vortex. Using this method, in two studies Barnes et al. \cite{barnes2018clockwise, barnes2018counterclockwise} used Taylor vortices to explore their interaction with a downstream NACA airfoil, examining vortex shedding and boundary layer effects. Similarly, Martinez-Muriel and Flores \cite{martinez2020analysis} investigated the effects of incoming Taylor and Lamb-Oseen vortices on the forces exerted on an airfoil at a Reynolds number of $1000$, and proposed a semi-empirical method to predict the response of the lift during vortex encounters. Similar approaches have also been used in studies that address other aspects of vortex-foil interactions. For example, Zhong et al.~\cite{zhong2023sparse} imposed Taylor vortices to investigate data-driven reconstruction of the wake dynamics during vortex–airfoil encounters, while Zhong et al.~\cite{zhong2025optimally} used the same framework to analyze the amplification of optimally time-dependent orthogonal modes in such interactions. While this strategy provides precise control over vortex properties and allows arbitrarily strong vortices to be considered, it is inherently limited to numerical studies, since imposing a predefined vortex within a flow field is not feasible in experiments. In addition, because the vortex is introduced directly into the computational domain, its consistency with the surrounding background flow may not always be fully ensured.
\par
A second type of vortex generation method, commonly used in experiments, involves either pitching or heaving a rigid body. Hufstedler and McKeon tested both pure heaving of a flat plate and pure pitching of an airfoil and observed that, although heaving can generate vortices, the resulting structures tend to be less compact than those generated through pitching \cite{hufstedler2019vortical}. Purely heaving-based approaches have also been explored by Qian et al. \cite{qian2022interaction}, who generated individual vortices through transient heaving motion of a hydrofoil. By varying the heaving velocity, they produced counter-clockwise and mostly two-dimensional vortices with different circulation magnitudes, but offered limited direct control over key vortex characteristics such as position and rotation orientation.
Peng and Gregory generated vortices by rapidly pitching an upstream airfoil and examined the resulting interaction modes with a downstream airfoil \cite{peng2015vortex}. Biler et al.~\cite{biler2021experimental} similarly used rapid pitching of a flat plate to generate vortex gusts and their unsteady interaction with an airfoil. Vadher and Babinsky \cite{vadher2024experimental} experimentally generated vortices of varying strength using linear pitching of an airfoil in a tow tank and found that the coherence of the resulting starting vortex depends strongly on the non-dimensional pitching distance.
Although pitching methods generate coherent vortices, a commonality across each of these prior works is that the stationary transverse position of the pitching body relative to the vortex produces a trailing wake that accompanies the vortex as it convects downstream. Hufstedler and McKeon \cite{hufstedler2019vortical} reported that this wake can significantly influence the downstream body, causing variations in the mean lift coefficient on the order of 0.1.
\par
Given the limitations of numerically introduced vortices and the constraints associated with commonly-used experimental generation methods, there remains a need for a vortex generation approach that is effective in both numerical and experimental contexts. To address this, we introduce a method for generating isolated vortex gusts using a symmetric airfoil undergoing combined heaving and pitching motions. This kinematic combination leaves the airfoil transversely offset from the dominant vortex trajectory, thus reducing additional wake influence beyond the dominant vortex itself. Direct numerical simulations (DNS) at low Reynolds numbers and water channel experiments with particle image velocimetry (PIV) at higher Reynolds numbers are performed to examine vortex formation and its subsequent interaction with a second airfoil positioned downstream of the vortex-generating airfoil. Vortex characteristics, including strength, size, and trajectory, are quantified, and comparisons between simulations and experiments are used to identify common trends. This method provides a promising basis for more controlled studies of isolated vortex gusts and their aerodynamic effects on downstream bodies.
\vfill

\section{Vortex Generation and Identification}
           \label{sec:generation}

\subsection{General configuration of vortex-generating airfoil}\label{ssec:airfoil_configuration}
Here we describe the generic vortex-generating system used in both simulations and experiments.  Details more specific to each approach are discussed in section III. Vortices are generated using a flapping symmetric NACA airfoil, with the pitch axis located at the quarter-chord. The chord length of the vortex-generating airfoil, denoted by $c$, is used as the reference length for nondimensionalization throughout this study unless otherwise specified.
\par
\begin{figure}[!htbp]
    \centering
    \includegraphics[width=0.7\linewidth]{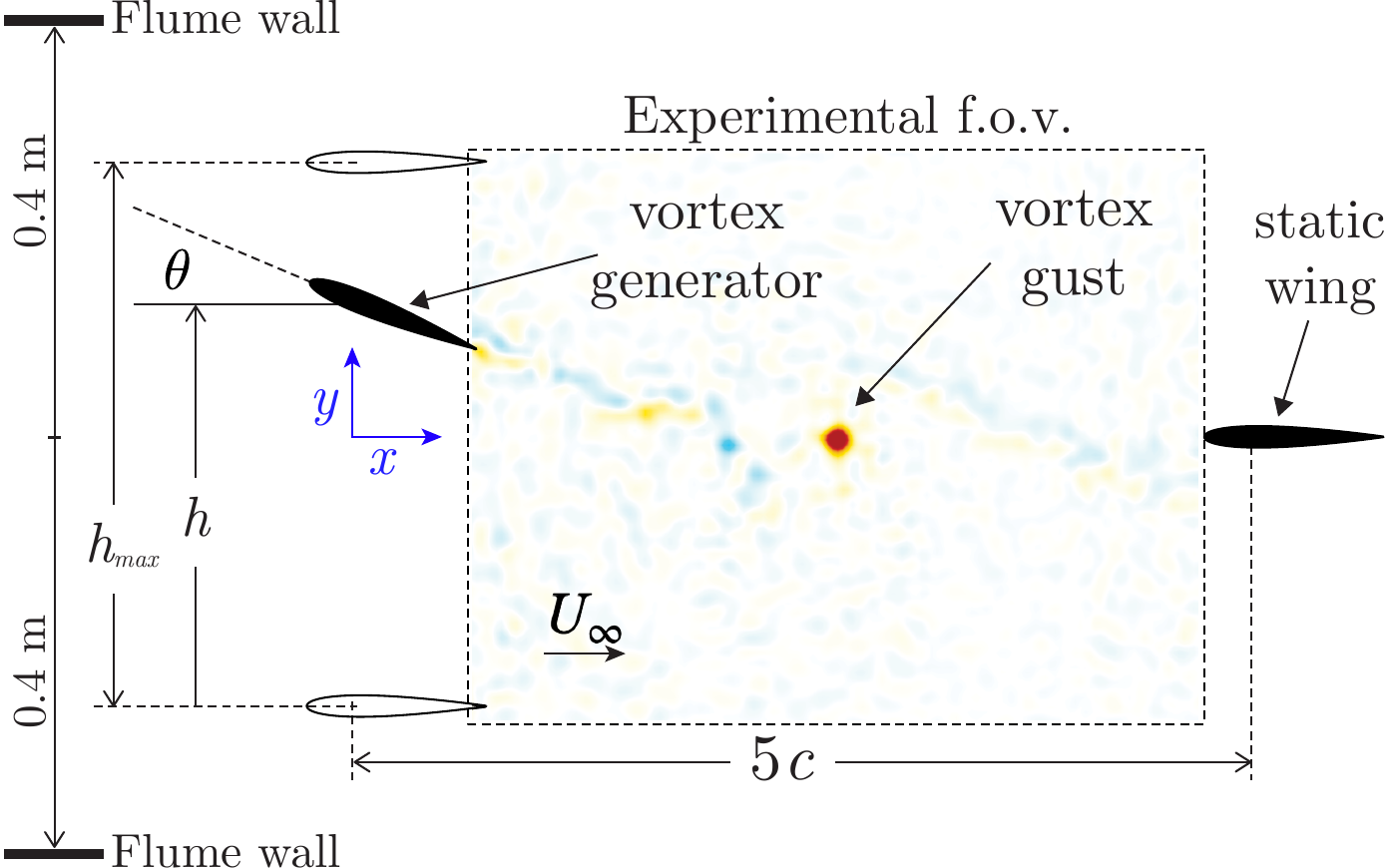}
    \caption{Schematic of the airfoil geometry and vortex-gust generation setup. The upstream airfoil undergoes prescribed heaving and pitching motions to generate vortex gusts, while a stationary downstream airfoil is positioned \(5c\) downstream. Flume walls are present only in the experimental configuration.}
    \label{fig:study_geometry}
\end{figure}

Figure~\ref{fig:study_geometry} shows a schematic of the two-airfoil configuration used in this study, consisting of an upstream vortex-generating airfoil and a downstream stationary airfoil that serves as a sensor for the incoming gust. The upstream airfoil undergoes prescribed heaving and pitching motions, while the downstream airfoil remains fixed.
\par
For both simulations and experiments, the streamwise distance between the quarter-chord points of the two airfoils is $5c$, and the upstream airfoil is initially positioned $1.333c$ below the downstream airfoil in the transverse direction. In the experiments, the heaving amplitude is limited to $h_{\max}=0.3\,\mathrm{m}$ by the gantry system, whereas no such constraint is imposed in the simulations. The experimental setup is further bounded by flume walls located $0.4\,\mathrm{m}$ from the flow centerline, while the simulations do not include confining wall boundaries. 

\subsection{Vortex generation profile}\label{ssec:vortex_profile}
Vortex generation is initiated through a rapid pitching event of the upstream airfoil, combined with simultaneous heaving motion.

To properly coordinate the pitch and heave components of the vortex-generating airfoil, we make use of the effective angle of attack, which directly influences the circulation around the airfoil. As shown by Simpson \cite{simpson2009experimental}, controlling the effective angle of attack also enables the suppression of higher harmonics that could otherwise lead to the formation of secondary, spurious vortices. The relationship between pitch, heave velocity, and effective angle of attack is given by
\begin{equation}
\alpha_{\mathrm{eff}}(t) = \theta(t) - \arctan{\frac{\dot{h}(t)}{U_\infty}},
\label{eqn:alphaeff}
\end{equation}
where \(\theta(t)\) is the pitch angle, \(\dot{h}(t)\) is the heave velocity, and \(U_\infty\) is the freestream velocity.

In this study, the flapping motion is modulated by prescribing the pitch angle and the effective angle of attack, while the heave motion is obtained numerically by inverting Eq.~\ref{eqn:alphaeff},
\begin{equation}
h(t) = \int_{0}^{t}U_\infty \tan{\left(\theta(\tau) - \alpha_{\mathrm{eff}}(\tau)\right)} d\tau.
\end{equation}
Although the motion involves both pitching and heaving, the effective angle of attack is designed to evolve in a manner similar to that observed in purely pitching cases, where the effective angle of attack reduces to the pitch angle. When heaving is included, the pitch angle and effective angle of attack can exhibit similar rates of change while differing in magnitude, allowing the combined motion to mimic the flow response of a pitching-only event. Having two degrees of freedom (pitch and heave) allows us both to define the effective angle of attack, \emph{and} to control the position of the airfoil wake relative to the shed vortex, thus providing  greater control over the resulting vortex characteristics. 
\par
To ensure smooth variation in the pitch angle and effective angle of attack, we introduce a transition function
\begin{equation}
r(t; t_{\mathrm{s}},t_{\mathrm{d}}) = \frac{1}{2} \left( 1 + \tanh\left(K\left(\frac{t-t_{\mathrm{s}}}{t_{\mathrm{d}}} - \frac{1}{2}\right)\right) \right),
\end{equation}
which provides a smooth transition in amplitude from $0$ to $1$, starting at time $t_{\mathrm{s}}$ and spanning a duration $t_{\mathrm{d}}$. The constant, $K=2 \ \mathrm{arctanh} (0.975)\approx4.369$, ensures that 97.5\% of the displacement occurs within time $t_d$. The pitch angle and the effective angle of attack are then expressed as combinations of transitions,
\begin{equation}\label{eqn:vortex_profile_theta}
\theta(t) = \theta_0 \, r(t; t_{\mathrm{s1}}, t_{\mathrm{d1}}) + \Delta \theta \, r(t; t_{\mathrm{s2}}, t_{\mathrm{d2}}) - (\theta_0 + \Delta \theta) \, r(t; t_{\mathrm{s3}}, t_{\mathrm{d3}}),
\end{equation}
\begin{equation}\label{eqn:vortex_profile_alpha}
\alpha_{\mathrm{eff}}(t) = \alpha_{\mathrm{eff},0} \, r(t; t_{\mathrm{s1}}, t_{\mathrm{d1}}) + \Delta \alpha_{\mathrm{eff}} \, r(t; t_{\mathrm{s2}}, t_{\mathrm{d2}}) - (\alpha_{\mathrm{eff},0} + \Delta \alpha_{\mathrm{eff}}) \, r(t; t_{\mathrm{s3}}, t_{\mathrm{d3}}).
\end{equation}
\par
\begin{figure}[h!]
\centering
	\begin{subfigure}{0.21\textwidth}
        \includegraphics[width=\textwidth]{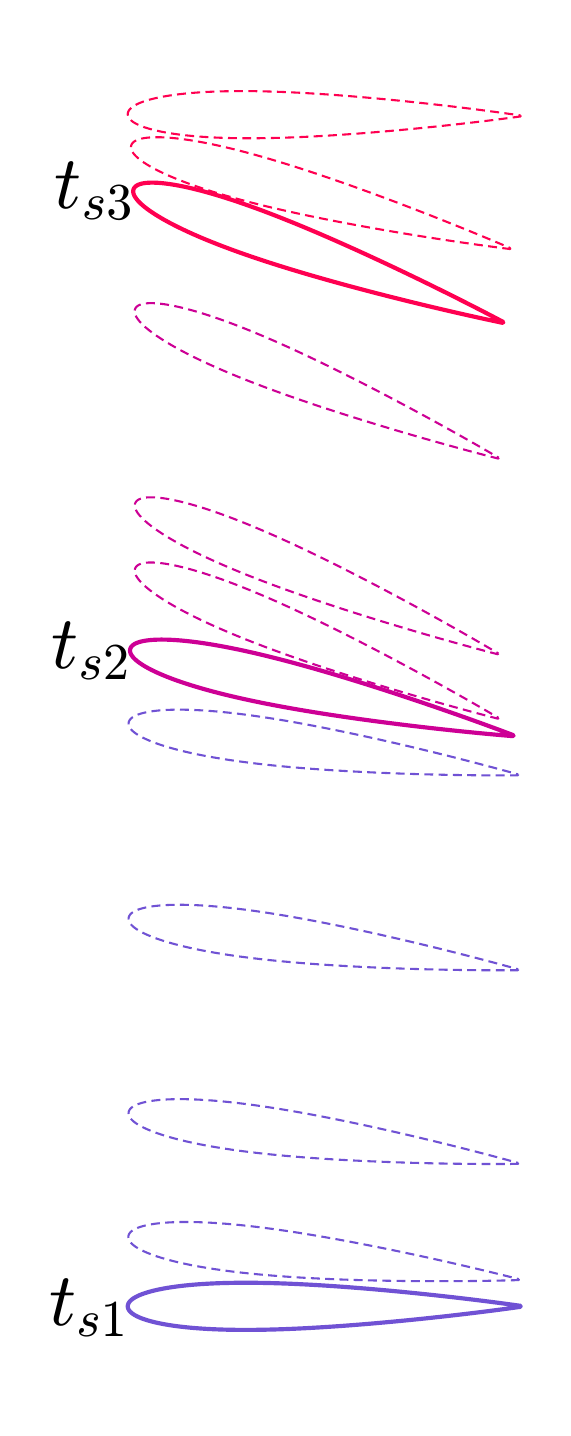}
        \caption{Foil position}
    \end{subfigure}
    \begin{subfigure}{0.71\textwidth}
        \includegraphics[width=\textwidth]{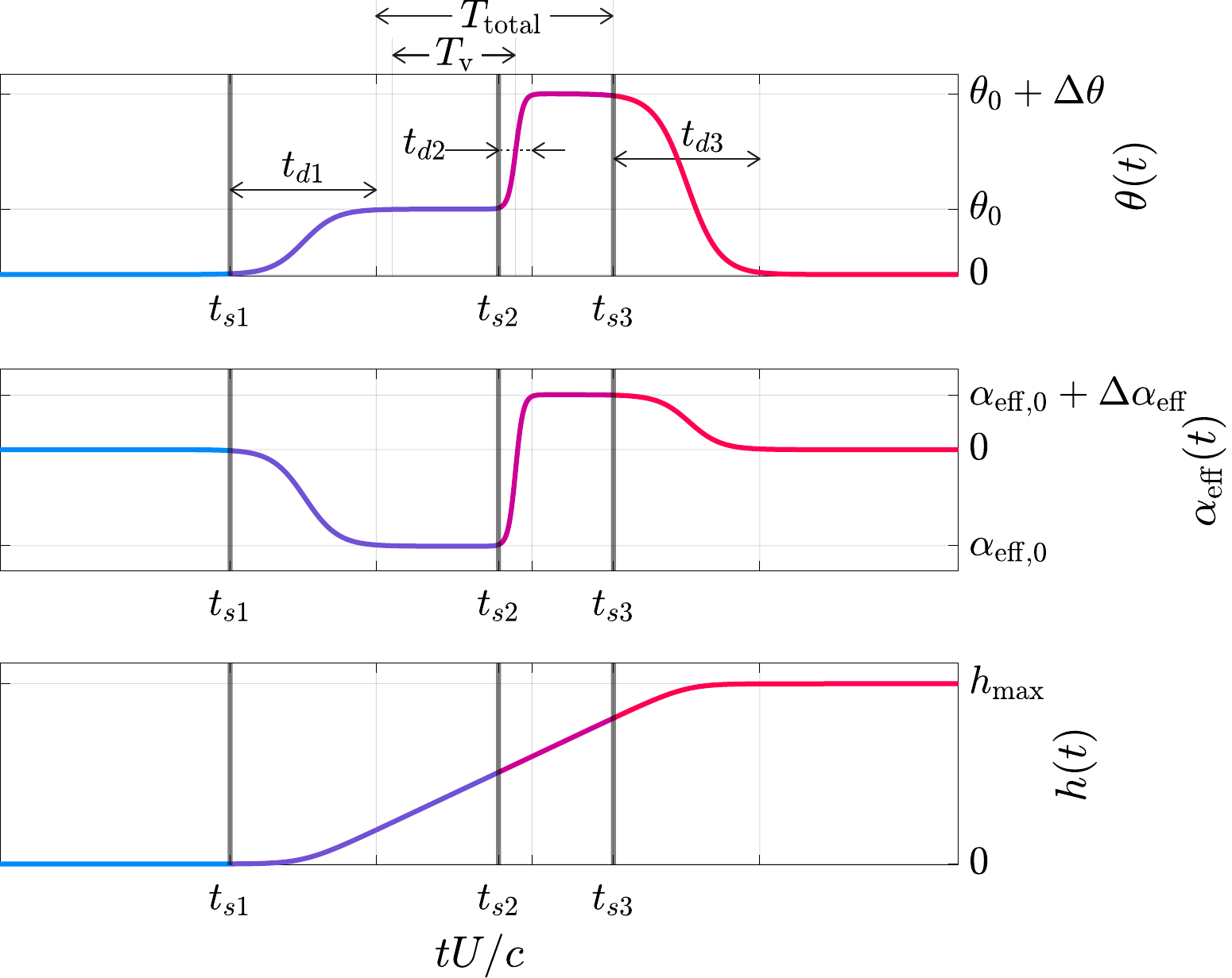}
        \caption{Pitch angle, effective angle of attack, and heave profiles.}
    \end{subfigure}
\caption{Foil kinematics during the vortex-generation maneuver. (a) Foil position colored according to the profiles shown in (b). (b) Kinematic profiles, with timing and geometric parameters defined in Eqs.~\ref{eqn:vortex_profile_theta} and \ref{eqn:vortex_profile_alpha}. Also shown are $T_\mathrm{total}$ and $T_\mathrm{v}$ from Eq.~\ref{eqn:tauv}.}
\label{fig:fli}	
\end{figure}

As illustrated in Figure \ref{fig:fli},  the foil starts at rest and then pitches up to achieve the first state at $\theta_0$ and $\alpha_{\mathrm{eff,0}}$.  Next, the second stage is a fast pitch-up or pitch-down maneuver, governed by $\Delta \theta$ and $\Delta \alpha_{\mathrm{eff}}$, which initiates shedding and roll-up of a trailing-edge vortex (TEV). Finally, the foil returns to zero pitch and zero angle of attack, but is now resting at a vertical position well above its start point.

\par
The primary vortex characteristics of interest in this study are the vortex \textit{strength}, \textit{direction of rotation}, and \textit{transverse position}. The parameters \( \Delta \theta \) and \( \Delta \alpha_{\mathrm{eff}} \) govern the magnitude of the rapid variation in angles and are expected to primarily influence the vortex strength, while their sign defines the rotation direction. For simplicity, these two parameters are set equal in this study. Prescribing different values would introduce additional heave motion, allowing part of the effective angle of attack variation to arise from the heaving velocity rather than from pitch alone. The baseline values, $\theta_0$ and $\alpha_{\mathrm{eff},0}$, primarily affect the foil's maximum heaving velocity and the total heaving distance traveled. Within the restricted range considered here, their influence on the resulting vortex characteristics is secondary compared to that of the rapid angular variations \( \Delta \theta \) and \( \Delta \alpha_{\mathrm{eff}} \).
\par
The time parameters $t_{\mathrm{s1}}$ and $t_{\mathrm{s3}}$ set the overall duration and influence the total heaving distance traveled by the foil. The parameter $t_{\mathrm{s2}}$ determines \textit{when} the vortex is produced and therefore controls its transverse location.
For convenience, we introduce the nondimensional timing parameter $\tau_\mathrm{v}$, which characterizes the relative placement of the vortex-generation onset within its admissible time window. Let $T_\mathrm{v}$ denote the interval between the chosen onset time and the earliest allowable onset time, and let $T_\mathrm{total}$ denote the width of the admissible onset interval, bounded by $t_{\mathrm{s1}}+t_{\mathrm{d1}}$ and $t_{\mathrm{s3}}-t_{\mathrm{d2}}$. The parameter $\tau_\mathrm{v}$ is then defined as
\begin{equation}\label{eqn:tauv}
    \tau_\mathrm{v}=\frac{ t_{\mathrm{s2}} - (t_{\mathrm{s1}}+t_{\mathrm{d1}})}{t_{\mathrm{s3}}-(t_{\mathrm{s1}}+t_{\mathrm{d1}}) - t_{\mathrm{d2}}} = \frac{T_\mathrm{v}}{T_\mathrm{total}}.
\end{equation}
The parameter $\tau_\mathrm{v}$ ranges from $0$ to $1$ and provides a convenient normalized measure of the onset timing, which is directly related to the transverse location of the generated vortex.
\par
We characterize the rapid pitching phase using a Strouhal number based on the duration $t_{\mathrm{d2}}$. Operating within an appropriate range of this parameter was found to promote the formation of compact, isolated vortices. The Strouhal number is defined as
\begin{equation}\label{eqn:st_td2}
    St_\mathrm{v}=\frac{|\dot{\alpha}_\mathrm{eff}|_{\mathrm{max}}c}{U_\infty}\approx \frac{|\Delta\alpha_\mathrm{eff}|Kc}{2t_\mathrm{d2}U_\infty},
\end{equation}
where \(|\dot{\alpha}_\mathrm{eff}|_\mathrm{max}\) is the maximum rate of change of the effective angle of attack during the rapid pitching phase. For a prescribed value of $St_\mathrm{v}$, the duration $t_{\mathrm{d2}}$ can be selected accordingly. In the experiments, $St_\mathrm{v}\approx0.655$ was used for all cases, as this value yielded coherent and isolated vortices over the range of conditions considered. In the simulations, coherent isolated vortices were obtained for $St_\mathrm{v}$ in the approximate range $1.63 \leq St_\mathrm{v} \leq 1.91$, which greatly differs from the experimental value. This discrepancy is likely associated with differences in Reynolds number and with the two-dimensional nature of the simulations compared with the three-dimensional experimental flow.
The remaining durations $t_{\mathrm{d1}}$ and $t_{\mathrm{d3}}$ do not correspond to the rapid pitching phase and are therefore not associated with a Strouhal number. Instead, these durations are chosen to be long enough such that any starting or ending vortices remain well separated from the primary vortex and have limited influence on the downstream airfoil.
\par
Together, these parameters enable the generation of vortices with controlled rotation direction, strength, and transverse position in a physically interpretable manner. 
A parametric investigation was conducted in which the change in effective angle of attack was varied over $-20^\circ\leq\Delta \alpha_\mathrm{eff}\leq20^\circ$, and the normalized onset parameter was varied within $0\leq\tau_\mathrm{v}\leq1$.
Each parameter combination is labeled according to whether it represents a simulation (S) or experiment (E), with the vortex rotation indicated as clockwise (CW) or counter-clockwise (CCW); e.g., S01\_CW denotes a clockwise vortex from a simulation, and E04\_CCW a counter-clockwise vortex from an experiment. The case numbering does not imply a one-to-one correspondence between simulation and experimental cases, although it may reflect comparable parametric trends. A full set of case-specific parameters is listed in Table~\ref{tab:detailed_param} in the Appendix.
Although the same parametric trends are examined in both simulations and experiments, the specific parameter values differ due to inherent differences between the numerical and experimental implementations, such as variations in Reynolds number. The numerical and experimental configurations are described in detail in \S~\ref{sec:methodology}.

\subsection{Vortex Identification and Classification}
To identify and characterize vortices, we adopt the $\Gamma_2$ criterion proposed by Graftieaux et al.~\cite{graftieaux2001combining}, which characterizes the local rotational organization of the velocity field and is invariant to uniform scaling of the velocity magnitude.
Although originally introduced for vortex identification in turbulent flows, Graftieaux et al. evaluated this criterion using a theoretical Lamb–Oseen vortex, finding that it provides a robust indicator of the vortex core and yields a nearly scale-independent threshold.
These properties make the $\Gamma_2$ criterion suitable for identifying vortices across the range of flow conditions considered in this study. In addition, the $\Gamma_2$ formulation provides a clear dimensionless threshold for defining vortex cores, enabling consistent identification across vortices of varying strengths.
\par
In both simulations and experiments, the value of \(\Gamma_2\) at a point \(\mathrm{P}\) is given by
\begin{equation}
\Gamma_2(\bm{x}_{\mathrm{P}}) = \frac{1}{|S_\mathrm{P}|}\int_{S_\mathrm{P}} \frac{\left[\left(\bm{x}-\bm{x}_{\mathrm{P}}\right) \times \left(\bm{u}- \tilde{\bm{u}}_\mathrm{P}\right)\right] \cdot \hat{z}}{||\bm{x}-\bm{x}_{\mathrm{P}}|| \, ||\bm{u}- \tilde{\bm{u}}_P||} \, dA,
\end{equation}
where \(S_\mathrm{P}\) is a small area surrounding point \(\mathrm{P}\), and
\begin{equation}
\tilde{\bm{u}}_\mathrm{P} = \frac{1}{|S_\mathrm{P}|} \int_{S_\mathrm{P}} \bm{u} \, dA
\end{equation}
is the average velocity within \(S_\mathrm{P}\).
Note that the $\Gamma_2$ calculation is performed on a two-dimensional velocity field. In the experiments, this field corresponds to the streamwise–transverse measurement plane located at the mid-span of the foil, whereas the simulations directly provide the required two-dimensional flow field. Further details of the numerical and experimental methodologies are provided in \S~\ref{sec:methodology}.
In the present experimental implementation, \(S_\mathrm{P}\) is taken as a small circular region of radius $0.2c$ surrounding point $\mathrm{P}$, while in simulations \(S_\mathrm{P}\) is taken as the union of all cells within 1.5$\Delta$ of point \(\mathrm{P}\), where $\Delta$ is the average cell size in the region. The vortex core \(S_{\mathrm{v}}\) is then defined as the region where \(|\Gamma_2| \ge 2/\pi\).
\par
Once the vortex region has been identified, the following characteristics are evaluated:

\paragraph{Vortex Strength}
The strength of each vortex is quantified by the circulation of its core, which in non-dimensional form is given by
\begin{equation}\label{eq:circulation}
    \Gamma_c^{*} = \frac{1}{U_{\infty} c} \int_{S_{\mathrm{v}}} \omega_z \, dA,
\end{equation}
where $\omega_z$ is the spanwise vorticity component.

\paragraph{Vortex Core Size}
The size of the vortex is represented by the equivalent circular radius of the core, defined as
\begin{equation}\label{eq:radius}
    r_c^{*} = \frac{\sqrt{S_{\mathrm{v}}/\pi}}{c}.
\end{equation}

\paragraph{Vortex Core Position}
The nondimensional coordinates of the vortex center, $(x_c^*,y_c^*)$, are defined using a vorticity-weighted centroid,
\begin{equation}\label{eq:vortex_position}
    x_c^{*} = \frac{\int_{S_{\mathrm{v}}} x \, \omega_z \, dA}{c\int_{S_{\mathrm{v}}} \, \omega_z \, dA}, \quad y_c^{*} = \frac{\int_{S_{\mathrm{v}}} y \, \omega_z \, dA}{c\int_{S_{\mathrm{v}}}  \, \omega_z \, dA},
\end{equation}
where \(x\) is referenced to the quarter-chord of the vortex generator, and \(y\) is referenced to the leading edge of the downstream airfoil.

\section{Numerical and Experimental Methods}\label{sec:methodology}
\subsection{Numerical Setup}
\label{sec:ge}

\subsubsection{Governing Equations and Solver}
\label{sec:gef}
The numerical simulations employ a NACA 0015 airfoil as the vortex generator and solve the incompressible Navier--Stokes equations at a chord-based Reynolds number of $Re=1\,000$. The equations are discretized using the open-source finite-volume libraries OpenFOAM, employing second-order spatial discretization schemes and a second-order backward temporal scheme. Pressure–velocity coupling is handled using the pressure implicit with splitting of operators (PISO) algorithm.
\par
The motion of the vortex-generating airfoil is accommodated using a mesh-morphing approach that solves for the updated location of each individual mesh point at each time-step. The mesh-motion equation combines a Laplacian term with a correction term accounting for solid-body rotation,
\begin{equation}
\nabla \cdot \left( 2 D_f \nabla \bm{d} \right)
+ \nabla \cdot \left( 
    D_f \left[ 
        \bm{n} \cdot \left( (\nabla \bm{d})^T - \nabla \bm{d} \right) 
        - \bm{n} \, \mathrm{tr}(\nabla \bm{d}) 
    \right]
\right) = 0,
\end{equation}
where $\bm{d}$ is the cell displacement vector, $D_f$ is the diffusivity controlling the rigidity and smoothness of the mesh deformation, and $\bm{n}$ is the normal vector of the cell faces. In the present implementation, the diffusivity is scaled inversely with the distance to the two airfoil surfaces, so that cells closer to either airfoil deform less and better retain their shape. The prescribed pitch and heave kinematics are imposed as boundary conditions on the surface of the vortex-generating airfoil in the mesh-motion solver, while zero displacement is enforced on the downstream airfoil and the outer domain boundaries. The mesh-motion methodology has been implemented and validated in previous studies \cite{ribeiro2021wake}.

\subsubsection{Computational Domain and Grid Independence}

\begin{figure}[!htbp]
     \centering
     \begin{subfigure}[b]{0.26\textwidth}
         \centering
         \includegraphics[width=\textwidth]{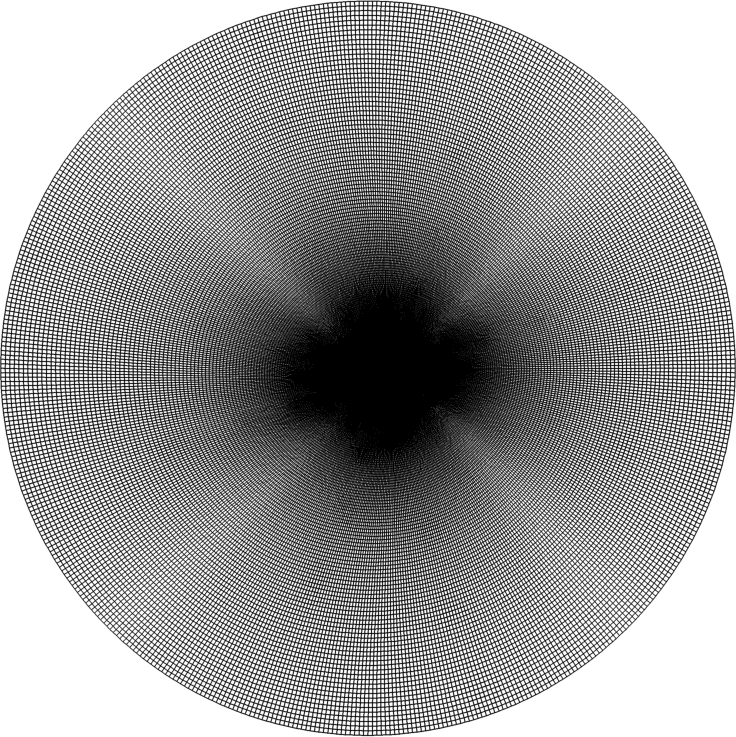}
         \caption{Overall domain}
     \end{subfigure}
     \hfill
     \begin{subfigure}[b]{0.34\textwidth}
         \centering
         \includegraphics[width=\textwidth]{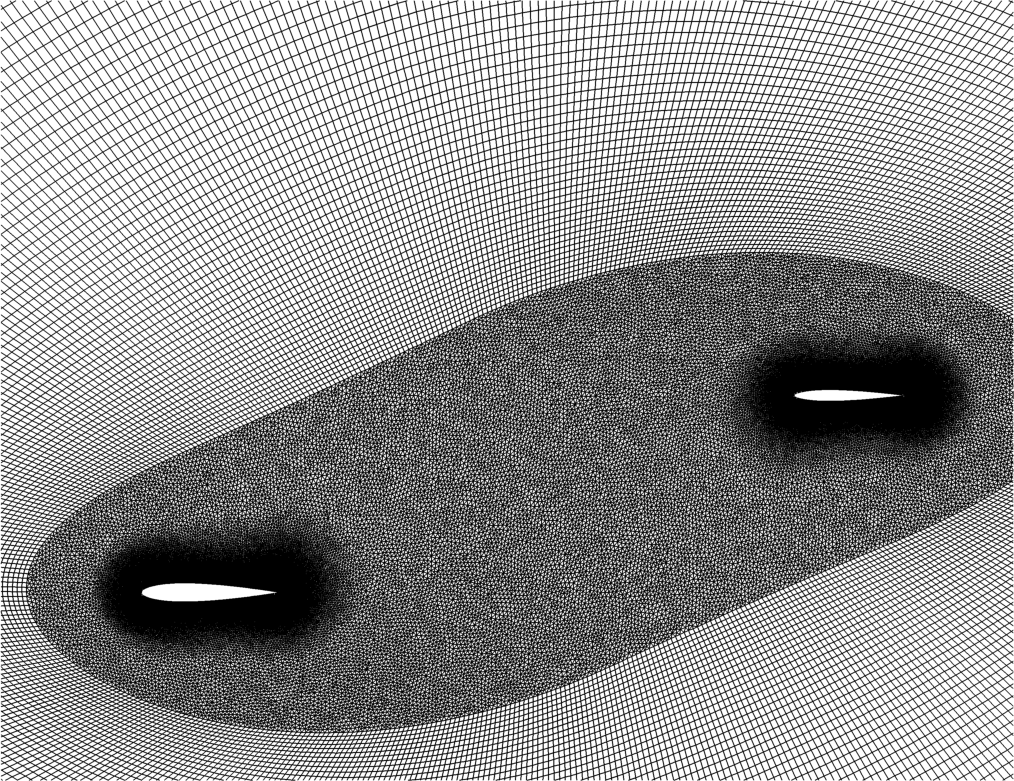}
         \caption{Intermediate mesh region}
     \end{subfigure}
     \hfill
     \begin{subfigure}[b]{0.34\textwidth}
         \centering
         \includegraphics[width=\textwidth]{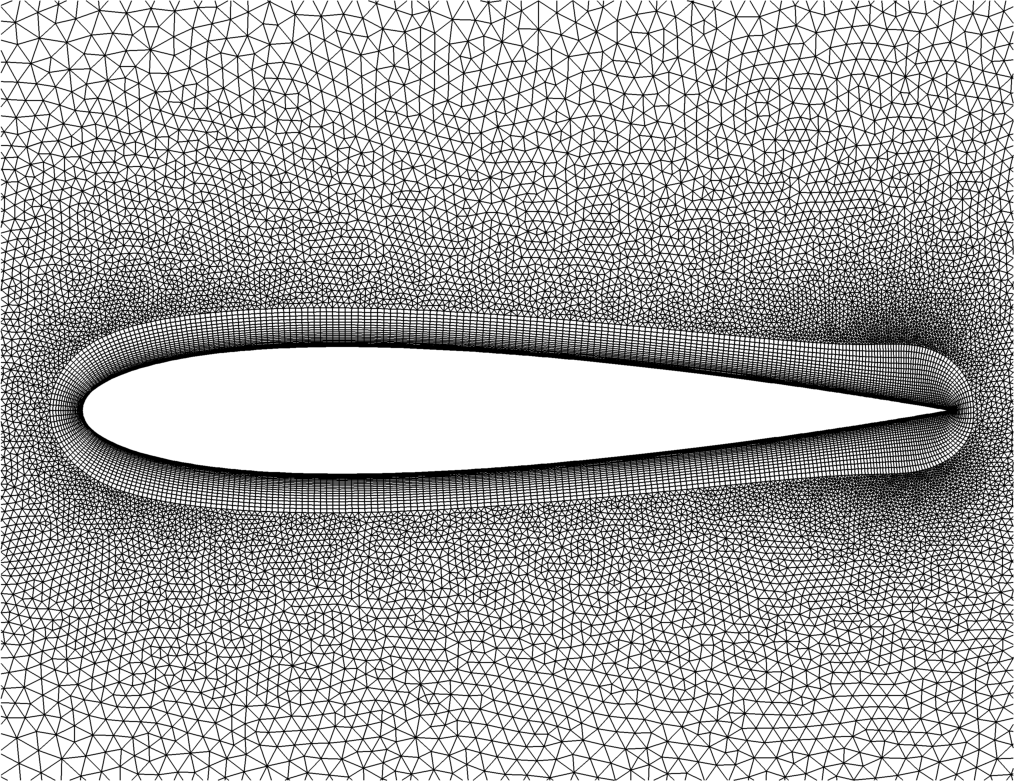}
         \caption{Near-airfoil mesh}
     \end{subfigure}
    \caption{Computational mesh layout used in the numerical simulations. (a) Overall computational domain. (b) Intermediate region where unstructured cells connect the near-airfoil meshes to the outer structured grid. (c) Body-fitted structured mesh near the upstream airfoil.}
    \label{fig:mesh}
\end{figure}

The computational domain, shown in Fig.~\ref{fig:mesh}(a), is circular, with the outer boundary located approximately \(50c\) from the vortex-generating foil. Standard incompressible inlet conditions are applied on the left half of the circular boundary, while outlet conditions are imposed on the right half. A no-slip condition is enforced on the surfaces of both airfoils. 
\par
The meshing strategy uses a predominantly structured grid throughout the domain, with an intermediate unstructured region connecting the structured blocks around the two airfoils. Body-fitted structured meshes are applied around each airfoil to accurately resolve boundary-layer separation and reattachment, as exemplified in Fig.~\ref{fig:mesh}(c). A separate structured mesh is also applied along the outer boundary to reduce numerical oscillations. The region between the body-fitted structured meshes around the airfoils and the outer structured mesh is discretized using unstructured cells, as shown in Fig.~\ref{fig:mesh}(b)-(c), to accommodate geometric complexity and mesh deformation, thereby avoiding excessive cell skewness or collapse.
\par

A grid independence study was conducted to ensure that the numerical results are not sensitive to grid resolution. Five mesh configurations with different resolution levels were tested, as summarized in Table~\ref{tab:grid}. Mesh3 serves as the baseline configuration. Compared with Mesh3, Mesh1 has a reduced resolution in the intermediate region between the airfoils, while Mesh4 increases the resolution in this region. Mesh2 reduces the resolution of the body-fitted meshes around both airfoils, whereas Mesh5 increases the resolution of the body-fitted meshes. For all meshes, simulations were performed using an identical flapping profile of the upstream airfoil. The lift coefficient of the downstream airfoil was monitored over a duration of 20 convective time units. The maximum and minimum lift coefficients, together with the circulation of the generated vortex averaged over a time interval during which the shed vortex convects toward the downstream airfoil, are reported in Table~\ref{tab:grid}.

\begin{table}[h!]
\caption{\small Mesh parameters and resulting aerodynamic quantities in the grid independence study. $N$: total number of cells, $\Delta_{\mathrm{wake}}$: approximate cell size between the airfoils, $N_{\theta}$: number of grid points along the top surface of the airfoil, $\Delta y_{1}$: first-layer mesh height around the upstream airfoil, $\Delta y_{2}$: first-layer mesh height around the downstream airfoil, $C_{l,\mathrm{max}}$: maximum lift coefficient of the downstream airfoil, $C_{l,\mathrm{min}}$: minimum lift coefficient of the downstream airfoil, $\overline{\Gamma}/(U_{\infty}c)$: mean non-dimensional circulation of the vortex core. The lift coefficient is nondimensionalized using the chord length of the downstream airfoil.}
\centering
\resizebox{\textwidth}{!}{%
\renewcommand{\arraystretch}{0.9}

\begin{tabular}{|*{9}{c|}}
\hline
 Mesh & $N$ & $\Delta_{\mathrm{wake}}/c$ & $N_{\theta}$ & $\Delta y_{1}/c$ & $\Delta y_{2}/c$ & $C_{l,\mathrm{max}}$ & $C_{l,\mathrm{min}}$ & $\overline{\Gamma}/(U_{\infty}c)$\\ \hline
 Mesh1 & $1.22 \times 10^{5}$ & $4.40 \times 10^{-2}$ & 200 & $1.00 \times 10^{-3}$ & $8.00 \times 10^{-4}$ & $6.11 \times 10^{-1}$ & $-9.66 \times 10^{-1}$ & $-0.554$\\ \hline
 Mesh2 & $1.57 \times 10^{5}$ & $3.04 \times 10^{-2}$ & 130 & $1.50 \times 10^{-3}$ & $1.20 \times 10^{-3}$  & $6.24 \times 10^{-1}$ & $-9.90 \times 10^{-1}$ & $-0.542$\\ \hline
 Mesh3 & $1.82 \times 10^{5}$ & $3.04 \times 10^{-2}$ & 200 & $1.00 \times 10^{-3}$ & $8.00 \times 10^{-4}$ & $6.20 \times 10^{-1}$ & $-9.92 \times 10^{-1}$ & $-0.543$ \\ \hline
 Mesh4 & $3.22 \times 10^{5}$ & $2.07 \times 10^{-2}$ & 200 & $1.00 \times 10^{-3}$ & $8.00 \times 10^{-4}$ & $6.25 \times 10^{-1}$ & $-1.01 \times 10^{0}$ & $-0.534$\\ \hline
 Mesh5 & $2.21 \times 10^{5}$ & $3.04 \times 10^{-2}$ & 300 & $6.00 \times 10^{-4}$ & $5.00 \times 10^{-4}$ & $6.18 \times 10^{-1}$ & $-9.89 \times 10^{-1}$ & $-0.542$\\ \hline
\end{tabular}%
}
\label{tab:grid}
\end{table}
\par
The results indicate that the maximum and minimum lift coefficients differ with variations of up to $1.2\%$ from Mesh2 to Mesh5. A similarly small variation is observed in the vortex circulation. These results suggest that the solution has largely converged with respect to grid resolution beyond Mesh3. Consequently, further mesh refinement yields only marginal changes in the computed quantities while greatly increasing the computational cost. Based on this analysis, Mesh3 is selected for the simulations presented in this study as it provides a suitable balance between accuracy and computational efficiency.

\subsection{Experimental setup}
\begin{figure}[h!]
    \centering
    \includegraphics[width=0.68\linewidth]{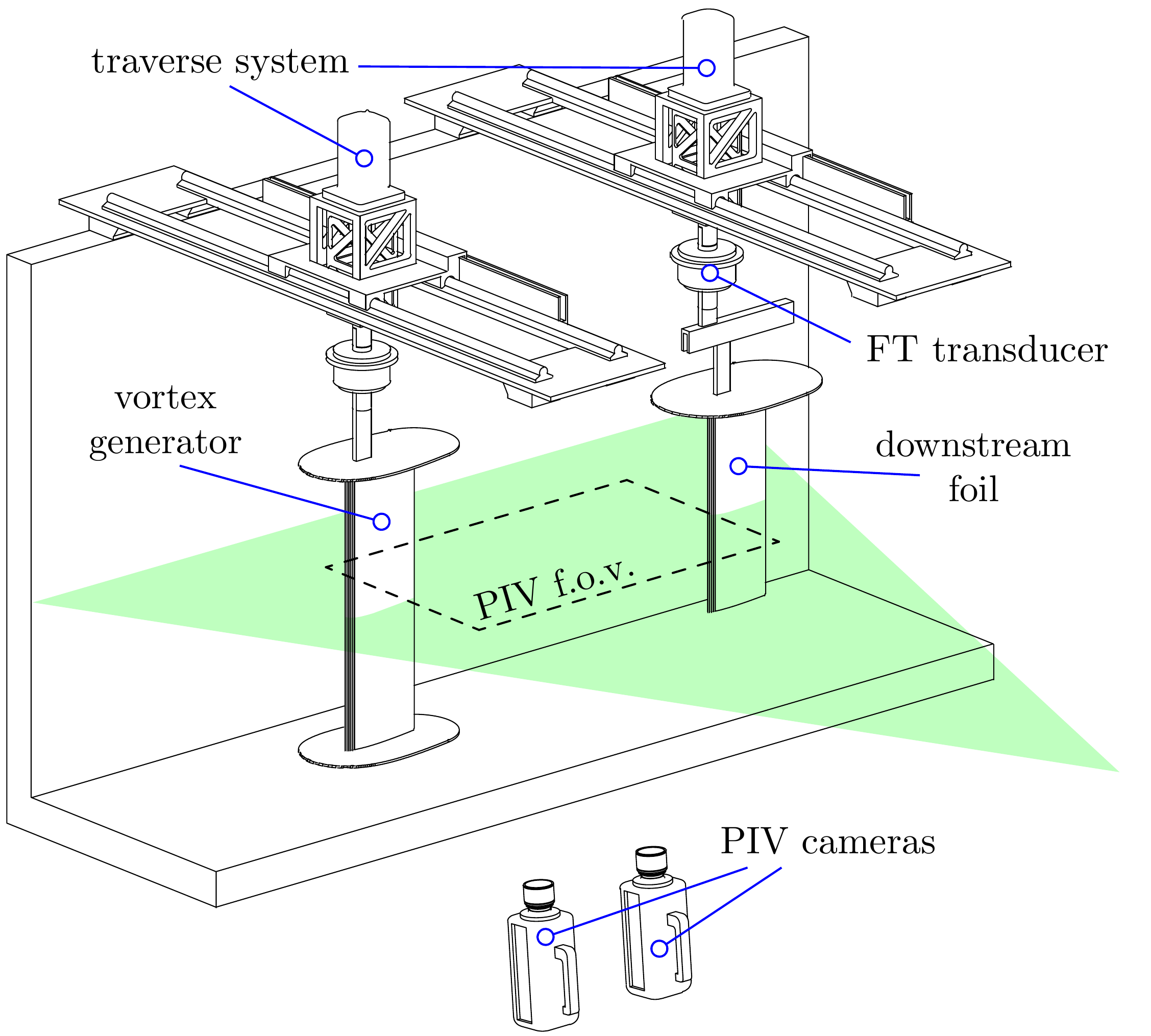}
    \caption{Experimental setup. The upstream symmetric airfoil serves as the vortex generator and is mounted on a traverse system. Two cameras arranged in a side-by-side configuration acquire PIV measurements in the measurement plane. A stationary downstream airfoil is positioned to measure the forces induced by the incoming vortex gust.}
    \label{fig:experimental_setup}
\end{figure}

Experiments were performed in an open-channel recirculating water flume at Brown University with test-section dimensions of $0.8\,\mathrm{m}\times0.6\,\mathrm{m}\times4\,\mathrm{m}$ (width, height, and length, respectively), as illustrated in Fig.~\ref{fig:experimental_setup}. The free-stream velocity was maintained at $U_\infty=0.35~\mathrm{m\,s^{-1}}$, measured upstream using an acoustic Doppler velocimeter (Nortek Vectrino). Three symmetric NACA-profile wings were used as vortex generators: a NACA~0015 with chord length $c=0.12\,\mathrm{m}$, and two NACA~0012 airfoils with $c=0.10\,\mathrm{m}$ and $c=0.08\,\mathrm{m}$, corresponding to chord-based Reynolds numbers of $42{,}000$, $35{,}000$, and $28{,}000$, respectively. Unless otherwise noted, the experimental flow-field snapshots presented in this work correspond to the $c=0.10\,\mathrm{m}$ cases. All wings had a span-length of $0.4\,\mathrm{m}$ and were equipped with end plates extending $0.5c$ from each tip to reduce three-dimensional effects. The prescribed pitch and heave motions of the vortex-generating airfoil were produced using a gantry traverse system mounted on the flume frame, driven by a Parker rotary servo for pitching and an AeroTech linear servo for heaving. The instantaneous pitch and heave positions were monitored independently using optical encoders (US Digital).
\par
Two-component, time-resolved PIV measurements were acquired in the stream\-wise–trans\-verse $xy$-plane at the mid-span location of the vortex-generating airfoil. Two high-speed cameras (Photron Fastcam Nova R2, $2048\times2048$ pixels) were positioned below the test section in a side-by-side arrangement, yielding a field of view of $0.4\,\mathrm{m}\times0.3\,\mathrm{m}$ in the $x$ and $y$ directions, respectively. The flow was illuminated using an Nd:YLF laser (Photonics Industries DM30), and image pairs were acquired at $6.66~\mathrm{Hz}$, with select cases acquired at $200~\mathrm{Hz}$. Velocity fields were computed using DaVis v10 (LaVision) and post-processed in MATLAB. A representative flow field acquired with this configuration is shown in Fig.~\ref{fig:study_geometry}.

\section{Results and Discussion}
\label{sec:result}
This results present the physics of the vortex generation process (\S~\ref{ssec:vortex_formation}) and an overview of the generated vortex characteristics (Table~\ref{tab:output_params} and \S~\ref{ssec:fundamental_vortex_characteristics}). Subsequently, a parametric study examining the influence of the pitch-heave profile parameters on resulting vortices is presented in \S~\ref{ssec:parametric_influence}, and the vortex-induced forces on the downstream airfoil are discussed in \S~\ref{ssec:vortex_induced_lift}. Finally, the implications of these trends are synthesized in \S~\ref{ssec:vortex_design}, where a framework is outlined for the controlled design of vortex gusts with prescribed properties.

\begin{table}[!htbp]
\centering
\caption{Summary of vortex characteristics in simulations and experiments. Experimental results shown correspond to $Re=35\,000$. Simulations are at $Re=1\,000$.}
\label{tab:output_params}

\setlength{\tabcolsep}{7 pt}
\renewcommand{\arraystretch}{1}

\begin{tabular}{|l|ccc||l|ccc|}
\hline
\multicolumn{4}{|c||}{\textbf{Simulations}} &
\multicolumn{4}{c|}{\textbf{Experiments}} \\
\hline
Case
& $\overline{y}^{*}$
& $\overline{\Gamma}_c^*$
& $\overline{r}_c^*$ 
&
Case
& $\overline{y}^{*}$
& $\overline{\Gamma}_c^{*}$ 
& $\overline{r}_c^*$ \\
\hline
S01\_CW  &  0.00 & -0.54 & 0.13 & E01\_CW   & -0.22 & -0.48 & 0.13 \\
S02\_CW  &  0.01 & -0.42 & 0.13 & E02\_CW   & -0.23 & -0.34 & 0.12 \\
S03\_CW  &  0.00 & -0.36 & 0.13 & E03\_CW   & -0.13 & -0.18 & 0.11 \\
S04\_CCW &  0.05 &  0.37 & 0.15 & E04\_CCW  & -0.24 &  0.21 & 0.11 \\
S05\_CCW &  0.07 &  0.41 & 0.14 & E05\_CCW  & -0.16 &  0.36 & 0.12 \\
S06\_CCW &  0.04 &  0.54 & 0.14 & E06\_CCW  & -0.18 &  0.46 & 0.13 \\
S07\_CW  & -0.47 & -0.55 & 0.13 & E07\_CW   & -0.72 & -0.46 & 0.13 \\
S08\_CW  & -0.24 & -0.54 & 0.13 & E08\_CW   & -0.40 & -0.47 & 0.13 \\
S10\_CW  &  0.24 & -0.55 & 0.14 & E09\_CW   & -0.04 & -0.48 & 0.13 \\
S11\_CW  &  0.48 & -0.56 & 0.14 & E10\_CW   &  0.29 & -0.47 & 0.13 \\
                 &    &    &    & E11\_CW   &  0.60 & -0.49 & 0.13 \\
                 &    &    &    & E12\_CW   & -1.03 & -0.44 & 0.13 \\
                 &    &    &    & E13\_CCW  & -1.02 &  0.46 & 0.13 \\
                 &    &    &    & E14\_CCW  &  0.65 &  0.47 & 0.13 \\
                 &    &    &    & E15\_CW   & -0.18 & -0.54 & 0.13 \\
                 &    &    &    & E16\_CCW  & -0.20 &  0.55 & 0.13 \\
\hline
\end{tabular}
\end{table}

\subsection{Vortex formation}\label{ssec:vortex_formation}

\begin{figure}[!h]
    \centering
    \includegraphics[width=\linewidth]{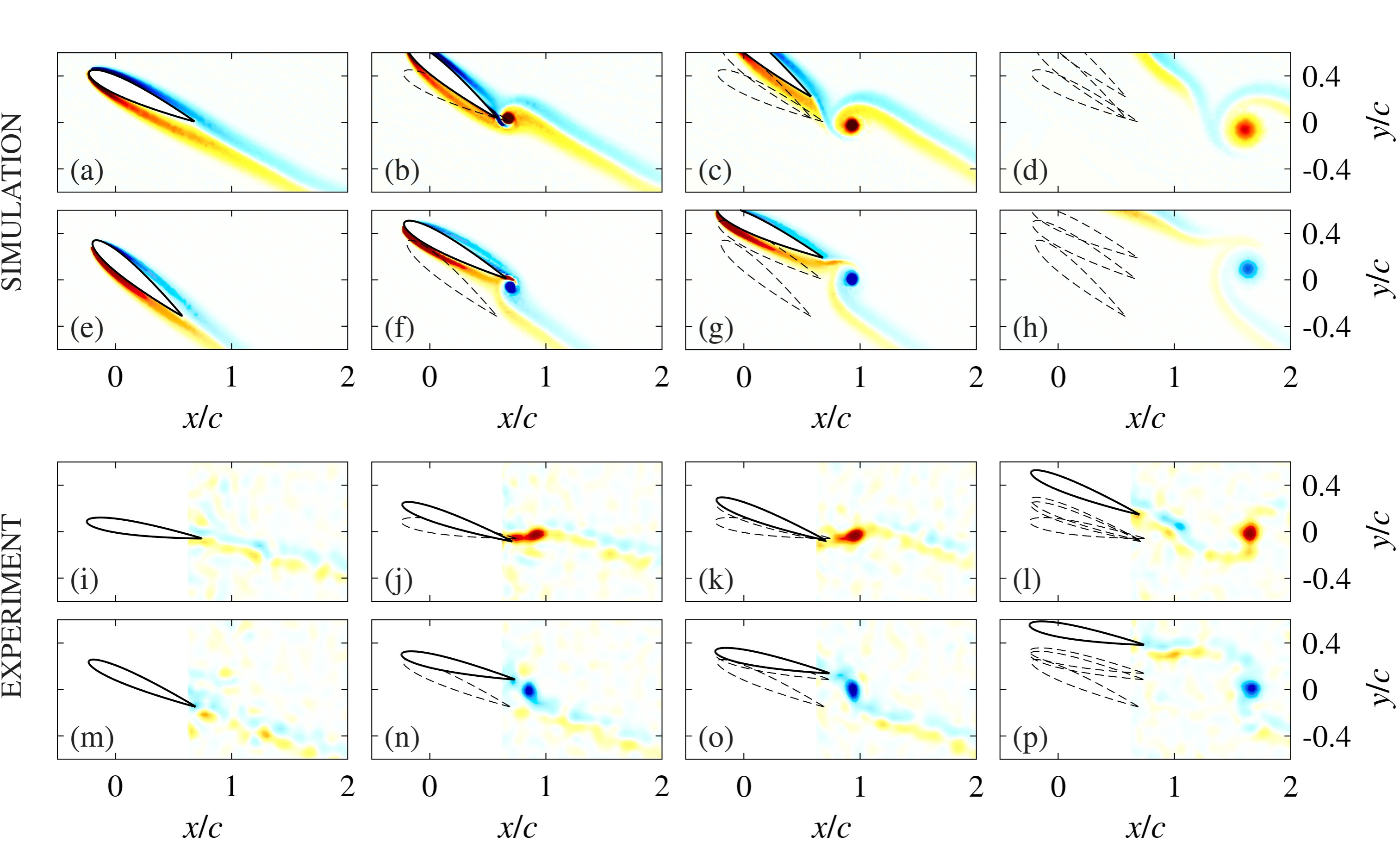}
    \caption{Time evolution of vorticity contours during the formation of counter-clockwise and clockwise vortices. Simulation results are shown for cases S06\_CCW (panels a--d) and S01\_CW (e--h), and experimental results for cases E06\_CCW (i--l) and E01\_CW (m--p).} \label{fig:vortex_formation_comparison}
\end{figure}

Figure~\ref{fig:vortex_formation_comparison} presents vorticity contours illustrating the vortex-formation process in both numerical simulations and experiments. Simulation results are shown in the first two rows and experiments in the third and fourth rows. The formation of CCW vortices is shown in panels~\ref{fig:vortex_formation_comparison}a--d for simulations and panels~\ref{fig:vortex_formation_comparison}i--l for experiments, while CW vortices are shown in panels~\ref{fig:vortex_formation_comparison}e--h for simulations and panels~\ref{fig:vortex_formation_comparison}m--p for experiments. From left to right, the four columns correspond to representative stages of the vortex-formation process: immediately prior to the rapid pitch maneuver, upon completion of the rapid pitch, during vortex shedding and roll-up, and after the vortex has fully detached and has begun advecting downstream.
\par
In the simulations, the shear layer over the upstream airfoil remains nearly steady prior to the onset of the rapid pitching phase, as both the pitch angle and the effective angle of attack are approximately constant during this period. Once the rapid pitching phase begins, the shear layer rapidly deforms and rolls up. A concentrated region of vorticity is subsequently shed from the trailing edge, forming a coherent vortex that propagates downstream, accompanied by a thin wake extending obliquely from the trailing edge. The sign of the rapid pitch maneuver determines the resulting trailing-edge vortex (TEV) rotation orientation: pitch-down motions ($\Delta \alpha_{\mathrm{eff}} < 0$) generate CW vortices, whereas pitch-up motions ($\Delta \alpha_{\mathrm{eff}} > 0$) generate CCW vortices. This behavior is consistent with the circulation sign induced by purely pitching airfoils in classical unsteady aerodynamics. 
\par
Although the experiments do not resolve the shear layer on the surface of the airfoil, the observed vorticity shedding follows a qualitatively similar sequence. This behavior closely resembles the TEV shedding process reported by Vadher and Babinsky \cite{vadher2024experimental} in their experimental vortex generation study. In the experiments, the vortex structures initially appear elongated during the shear-layer roll-up (e.g., panels (j)–(k) for the CCW case and panels (n)–(o) for the CW case in Fig.~\ref{fig:vortex_formation_comparison}), a feature also observed by Vadher and Babinsky. In contrast, the vortices in the simulations appear less elongated during the roll-up process. This difference is likely associated with the much smaller $St_\mathrm{v}$ used in the experiments, which corresponds to a lower nondimensional pitch rate during the rapid pitching phase and consequently faster TEV formation. These observations demonstrate that the rapid pitching motion of the upstream airfoil produces compact, well-defined vortices that propagate downstream with limited continuous wake interference.

\subsection{Fundamental vortex characteristics}\label{ssec:fundamental_vortex_characteristics}

\subsubsection{Vortex trajectory, circulation and core radius}
\begin{figure}[!htbp]
\centering
	\begin{subfigure}{0.328\textwidth}
        \includegraphics[width=0.9\textwidth]{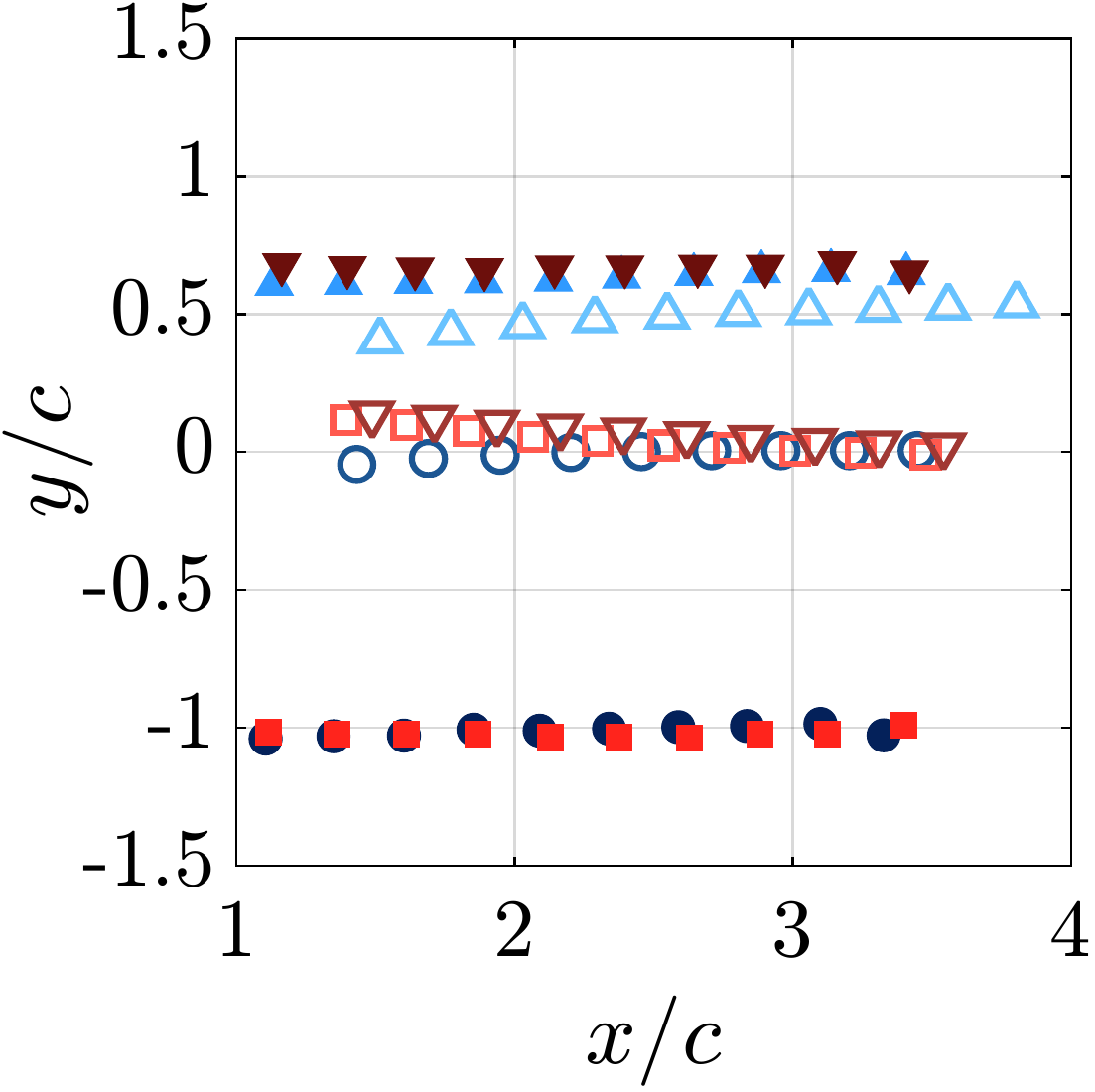}
        \caption{Vortex trajectory.}\label{fig:vtx_trajectory}	
    \end{subfigure}
    \begin{subfigure}{0.328\textwidth}
        \includegraphics[width=0.9\textwidth]{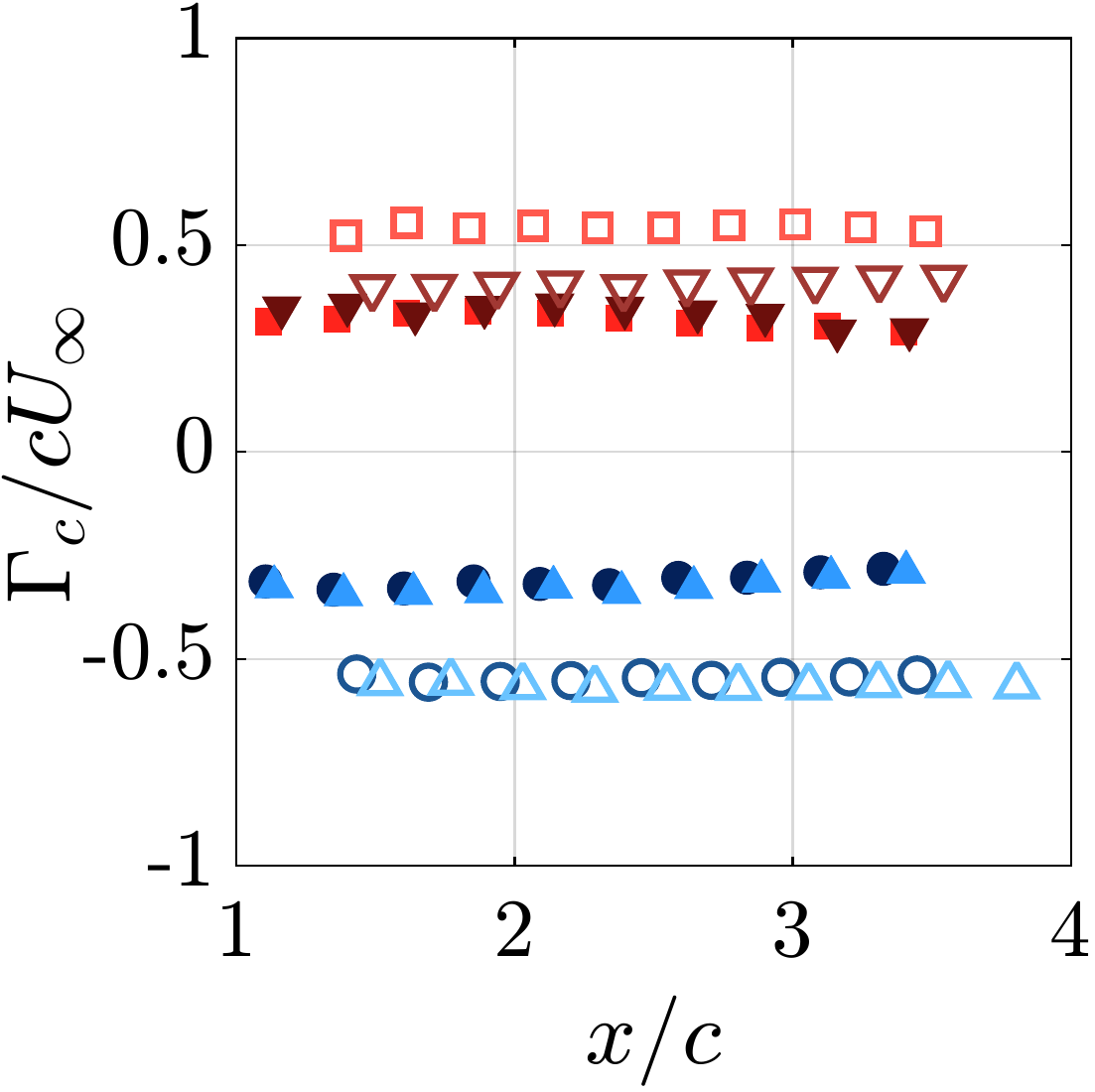}
        \caption{Vortex circulation.}\label{fig:vtx_strength}	
    \end{subfigure}
    \begin{subfigure}{0.328\textwidth}
        \includegraphics[width=0.9\textwidth]{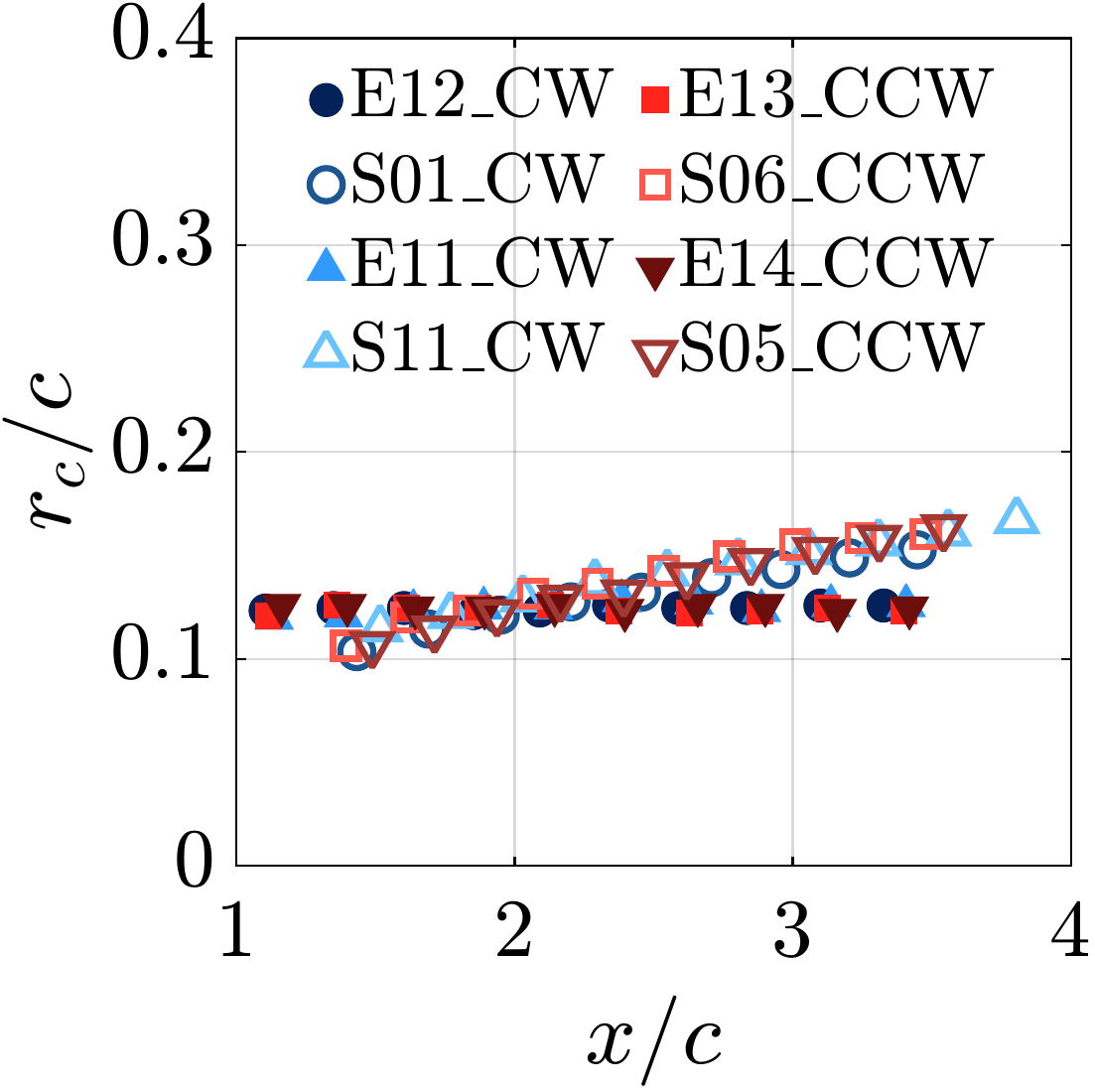}
        \caption{Vortex core radius.}\label{fig:vtx_radius}	
    \end{subfigure}
\caption{Trajectory, circulation, and core radius of representative cases from simulations and experiments (four cases each).}
\label{fig:trajectory_strength_radius}	
\end{figure}

The average values of $y$-position, circulation, and core radius resulting from the vortex characterization are shown in Table~\ref{tab:output_params}.
Additionally, Figure~\ref{fig:trajectory_strength_radius} summarizes the primary vortex trajectory, circulation, and core radius during convection for several representative simulation and experimental cases.

Despite the large transverse velocity of the flapping airfoil during vortex shedding, the vortices advect primarily in the streamwise direction with only small transverse displacement until they approach the downstream airfoil, where they either directly impinge on it or are deflected. The average streamwise advection velocity of the vortex is close to the freestream velocity, remaining within approximately $1\%$ of $U_\infty$ in experiments. For simulations a CW vortex advects at the freestream velocity whereas the CCW vortcies have an average advection velocity of $0.90-0.93U_{\infty}$. The transverse advection velocity remains small, not exceeding $0.05U_\infty$ in experiments and $0.07U_\infty$ in simulations. The larger deviation in the CCW simulations likely reflect differences in the vortex generation process, which can impart positive $x$-momentum (CCW) or negative $x$-momentum (CW) during the rapid pitch motion. In experiments, the rapid pitch rate is lower, and thus will inject smaller amplitudes of $x$-momentum on the vortex. Despite these factors, the vortex trajectories show that the generated vortices are well suited for studying vortex–airfoil interactions in which the vortex approaches primarily along the streamwise direction.
\par
Throughout this advection process, the vortex circulation remains approximately constant, indicating that the vortices remain well defined before interacting with the downstream airfoil. However, as observed from Table~\ref{tab:output_params} and Figure~\ref{fig:exp_vortex_strength}, vortices in simulations display slightly higher circulation than experiments, something observed for all $Re$ explored in experiments. The evolution of the vortex core radius differs between simulations and experiments. The simulation cases show a gradual increase in radius during convection, occurring at a comparable rate across the different cases, whereas the experimental cases remain nearly constant. This difference is likely associated with the lower Reynolds number used in the simulations, which enhances viscous diffusion, as well as with the two-dimensional nature of the numerical model, which inhibits vortex stretching.
\subsubsection{Vortex-induced velocity profile}
To further assess the coherence of the generated vortices, induced velocity profiles are extracted and compared with the theoretical Lamb--Oseen vortex model (LO-model). The center of each vortex is first identified, after which the $y$-velocity is extracted along a streamwise slice across the vortex center and the $x$-velocity along a transverse slice through the same point. The azimuthal velocity profile of the Lamb--Oseen vortex model is given by
\[
v_\theta(r) = \frac{\Gamma_{\mathrm{total}}}{2\pi r} \left(1 - e^{-kr^2}\right),
\]
where $\Gamma_{\mathrm{total}}$ denotes the total circulation and $k$ is a parameter controlling the radial spread of the vortex.
It should be noted that $\Gamma_{\mathrm{total}}$ differs from the circulation of the vortex core obtained from the $\Gamma_2$-based identification procedure. For an ideal Lamb--Oseen vortex, application of the $\Gamma_2$ criterion identifies a core region approximately bounded by the radius at which the azimuthal velocity reaches its maximum \cite{graftieaux2001combining}, corresponding to $r \approx 1.121/\sqrt{k}$.
This region contains a circulation of approximately $\Gamma_c \approx 0.715\,\Gamma_{\mathrm{total}}$.
\par
\begin{figure}[!htbp]
\centering
	\begin{subfigure}{0.49\textwidth}
        \includegraphics[width=0.76\textwidth]{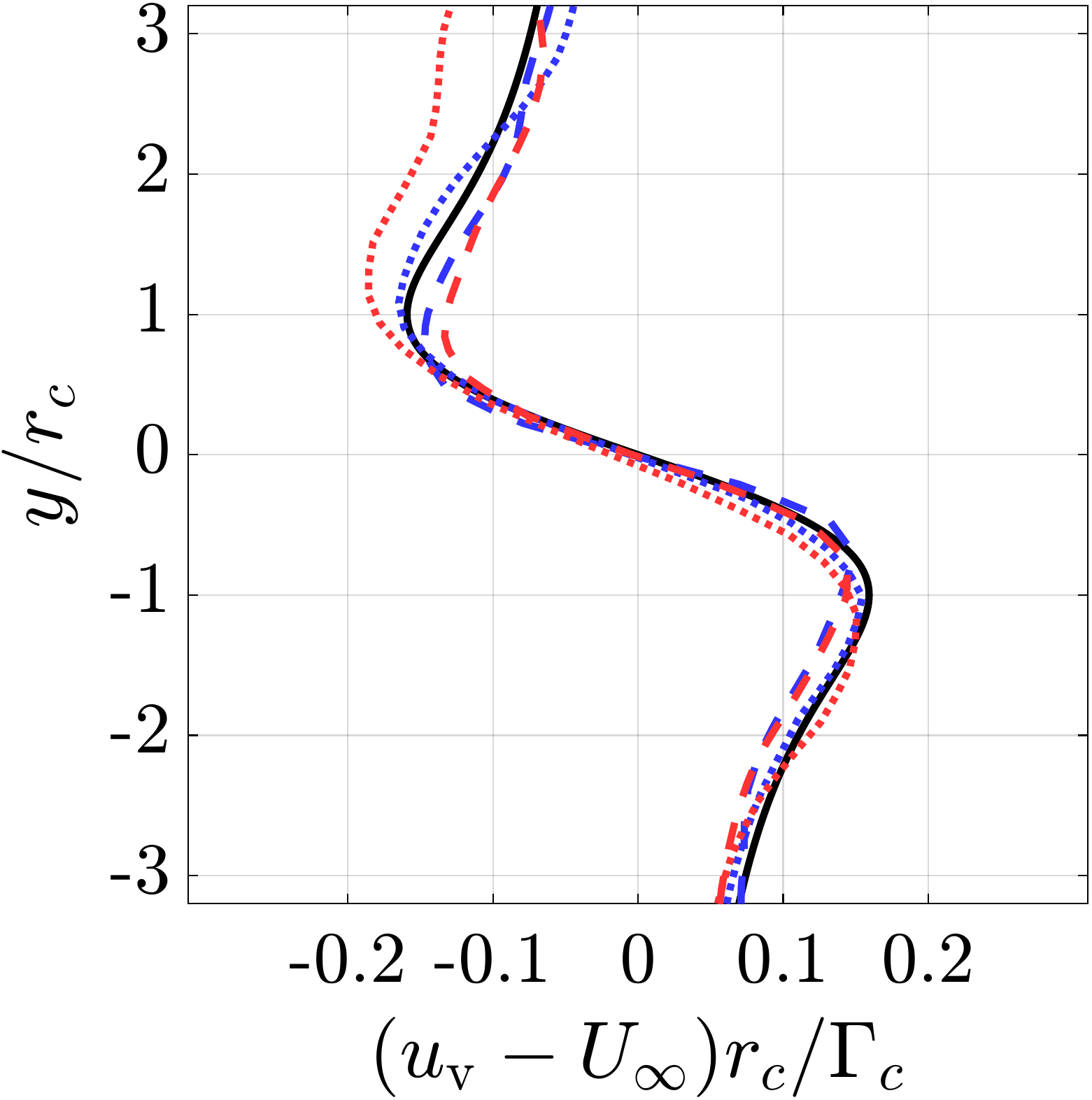}
        \caption{Streamwise gust-induced velocity profile.}
    \end{subfigure}
    \begin{subfigure}{0.49\textwidth}
        \includegraphics[width=0.8\textwidth]{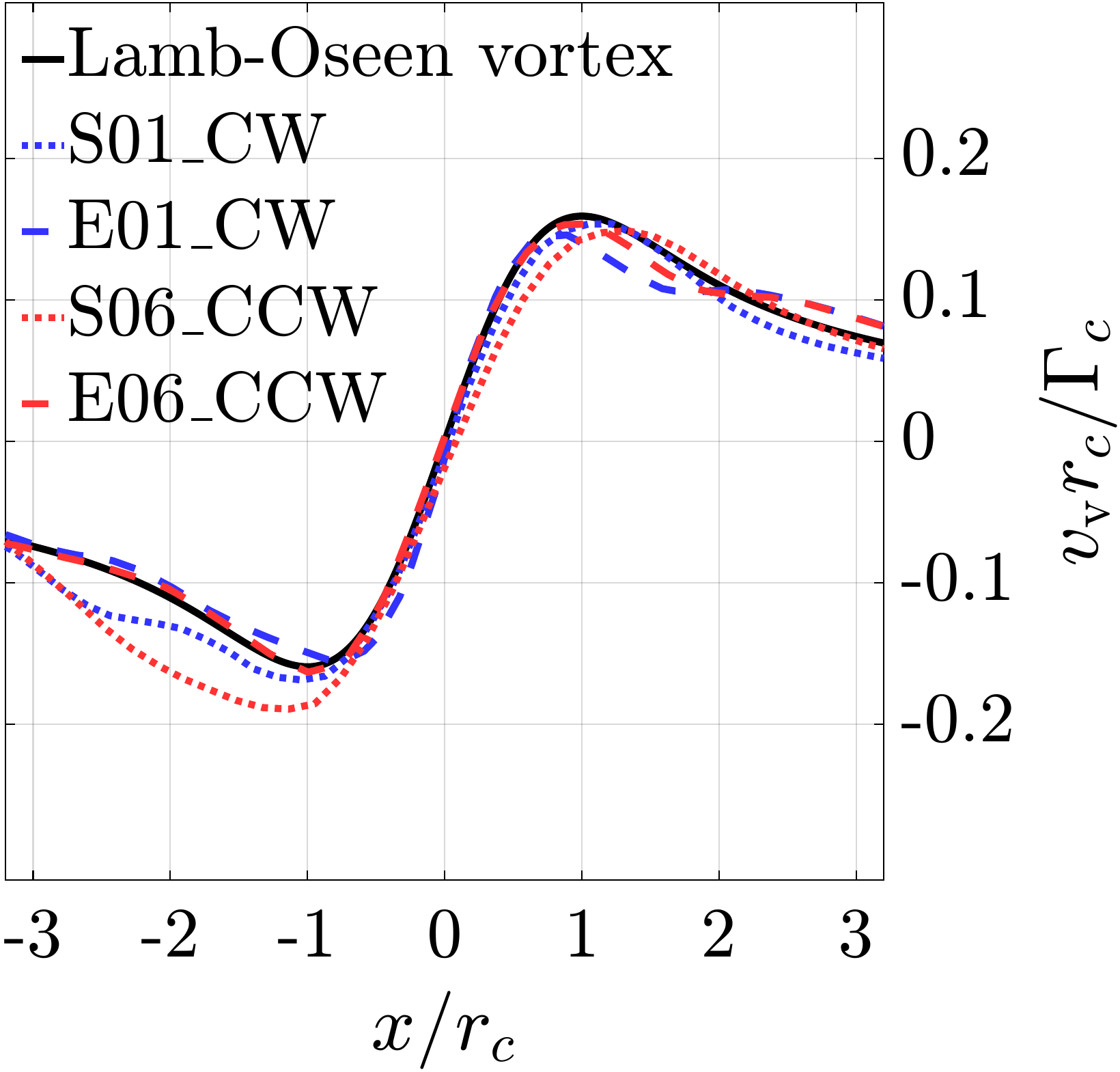}
        \caption{Transverse gust-induced velocity profile.}
    \end{subfigure}
\caption{Non-dimensional velocity profile across the vortex center. Simulation and experimental results are compared with the Lamb-Oseen vortex model. (a) shows the variation of streamwise gust velocity $u_\mathrm{v}$ along $y/r_c$, and (b) shows transverse gust velocity $v_\mathrm{v}$ along $x/r_c$. Both velocities are normalized by the vortex-core circulation $\Gamma_c$.}
\label{fig:vortex_velocity_profile}	
\end{figure}

To compare the measured velocity fields with the LO-model, representative vortices with similar strength are selected, and the model parameters are obtained through a nonlinear least-squares fit to the simulation and experimental data. The resulting velocity profiles are shown in Fig.~\ref{fig:vortex_velocity_profile}. To enable meaningful comparisons across simulations, experiments, and the LO-model, the profiles are rescaled using the circulation and core radius of the identified vortices obtained. The measured profiles exhibit excellent agreement with the LO-model, confirming the compactness and coherence of the generated vortices. 

\subsection{Parametric influence on vortex behavior}\label{ssec:parametric_influence}
\begin{figure}[!h]
    \centering
    \includegraphics[width=\textwidth]{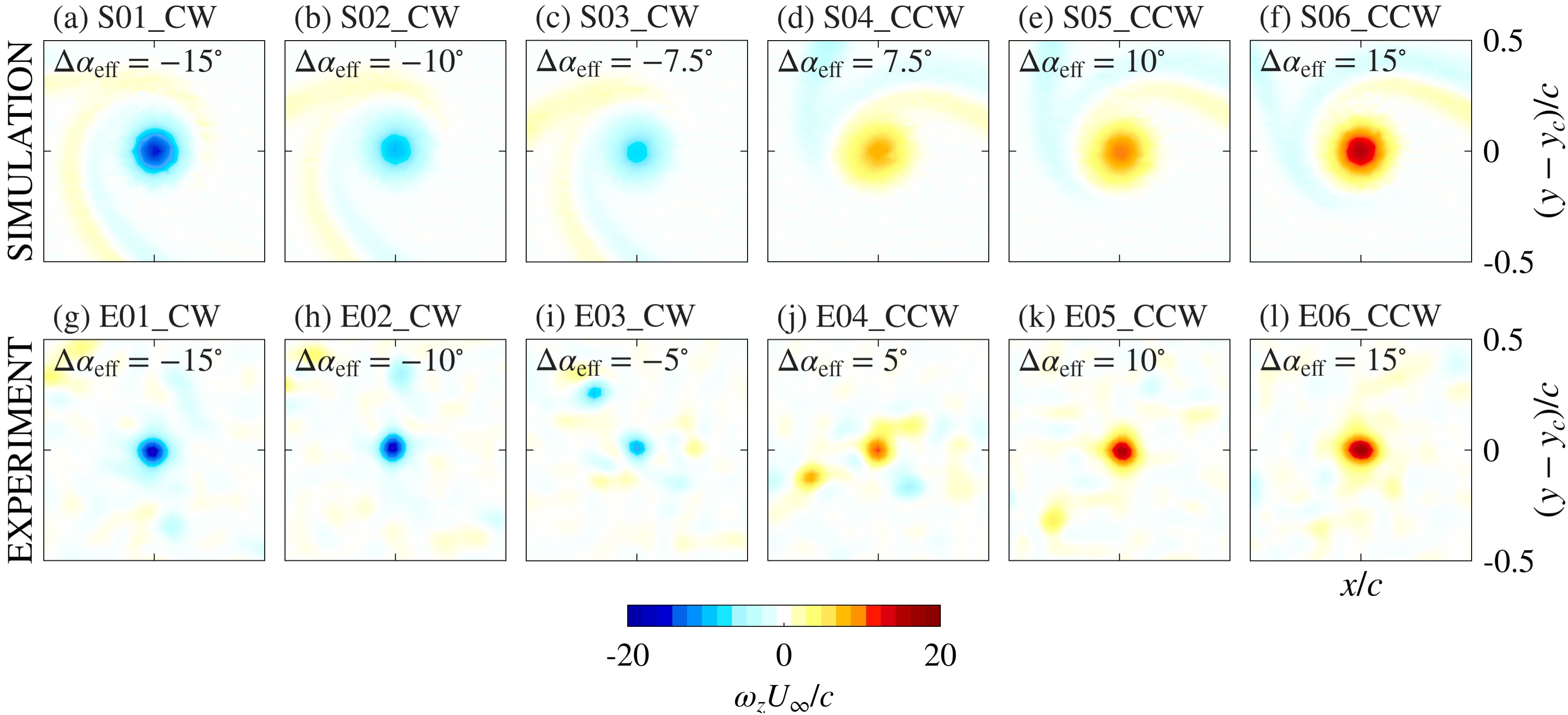}
    \caption{Vorticity contour snapshots illustrating the variation of vortex strength with the rapid angle of attack maneuver amplitude $\Delta\alpha_{\mathrm{eff}}$.}

    \label{fig:exp_vortex_strength}
\end{figure}

\begin{figure}[!h]
    \centering
    \includegraphics[width=\textwidth]{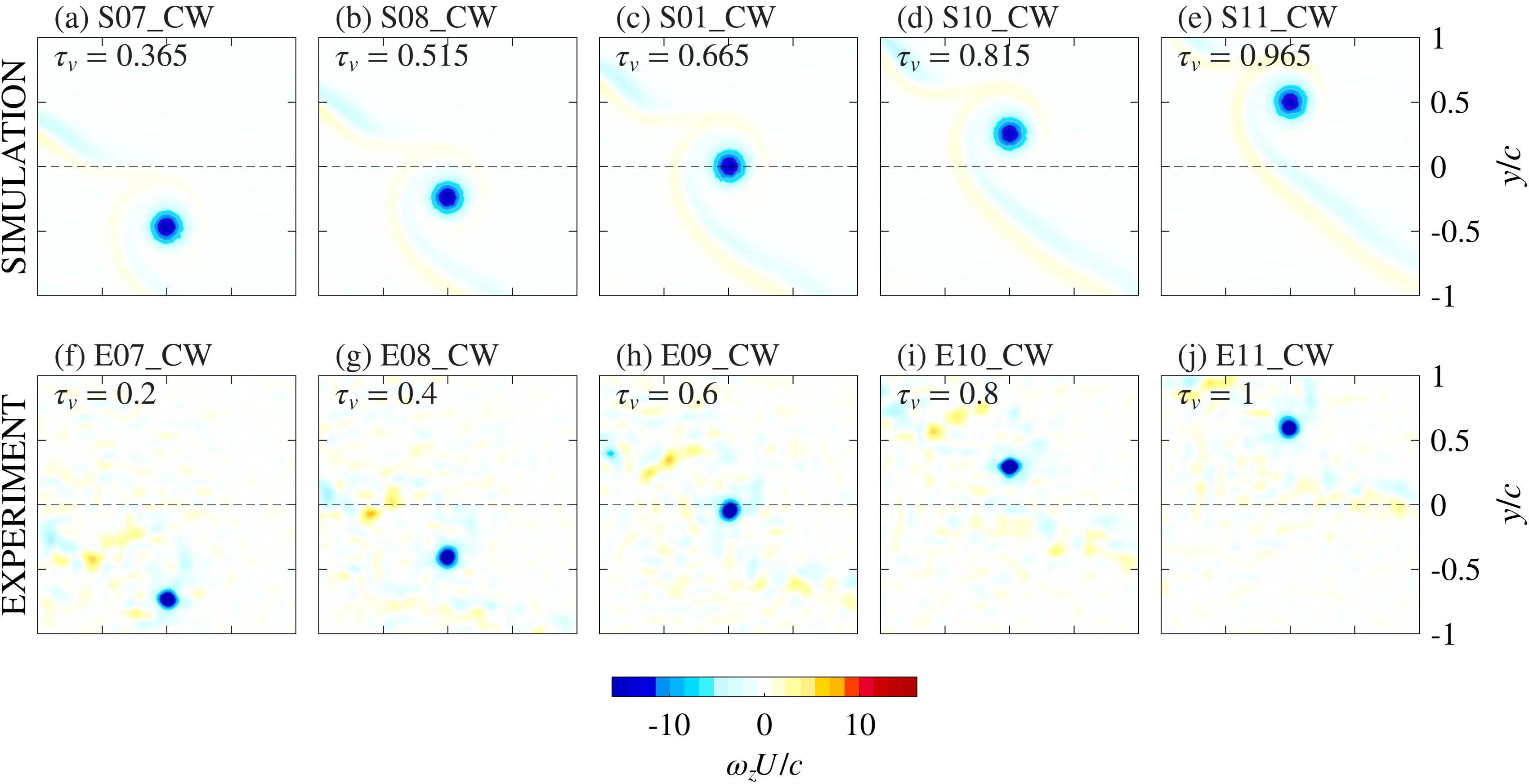}
    \caption{Vorticity contour snapshots illustrating the variation of vertical vortex position with the shedding instance parameter $\tau_{\mathrm{v}}$. In these cases, $\tau_{\mathrm{v}}$ (and the corresponding shedding time $t_{\mathrm{s2}}$) is varied, while all other flapping profile parameters are held constant.}
    \label{fig:exp_vortex_position}
\end{figure}

\begin{figure}[!htbp]
\centering
	\begin{subfigure}{0.48\textwidth}
        \includegraphics[width=0.8\textwidth]{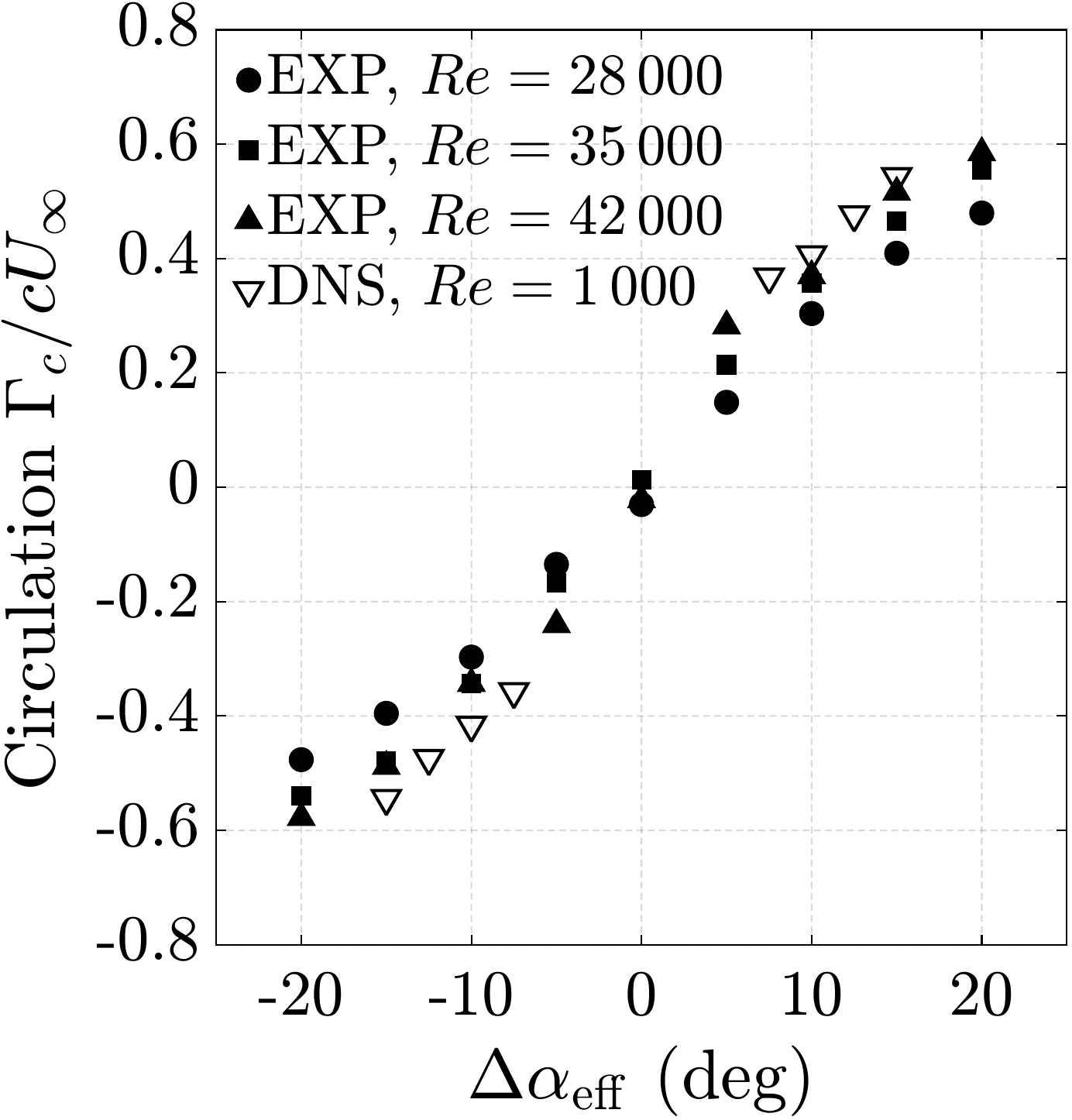}
        \caption{Vortex strength variation.}\label{fig:alpha_vs_circulation}
    \end{subfigure}
    \begin{subfigure}{0.48\textwidth}
        \includegraphics[width=0.8\textwidth]{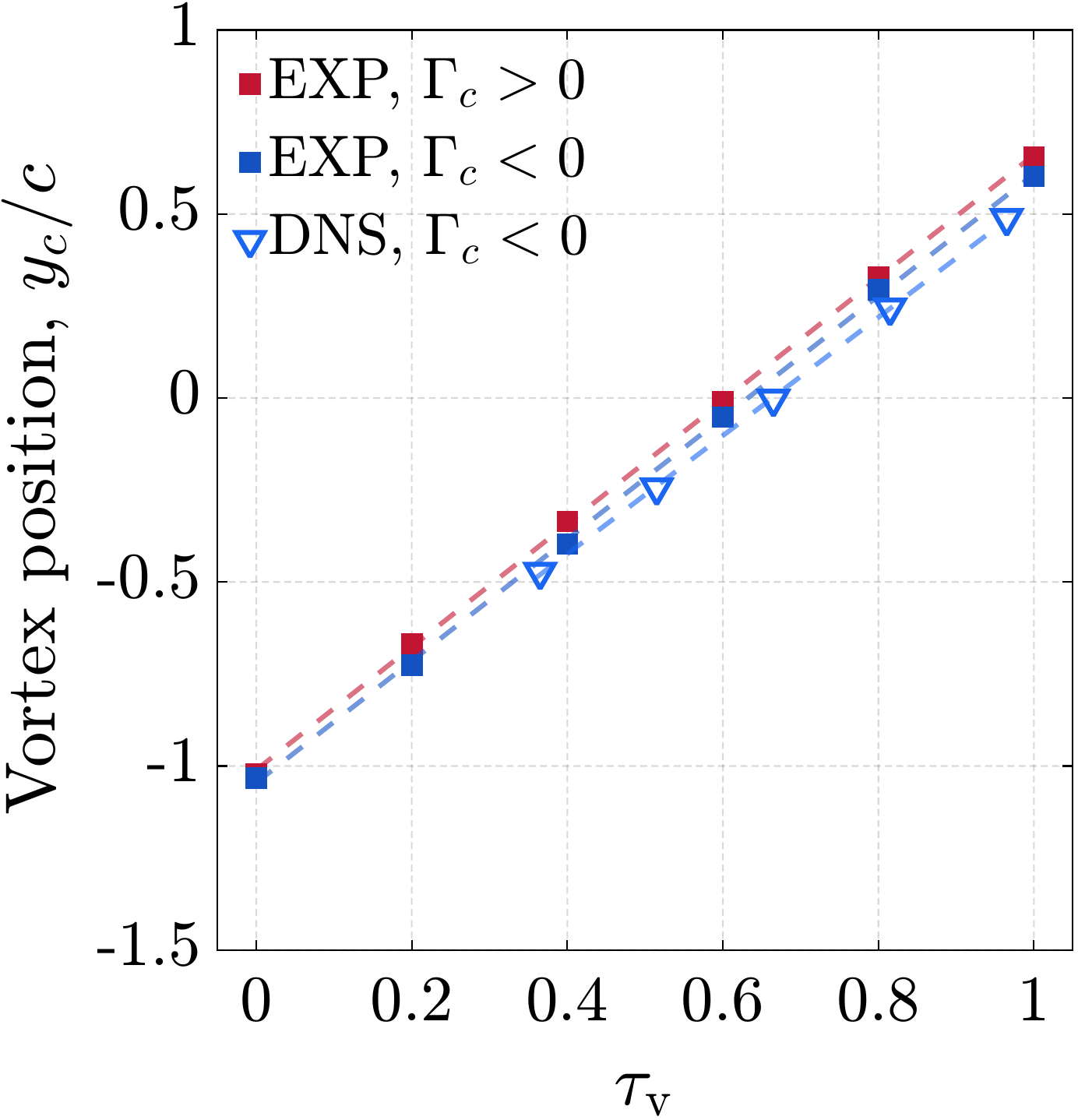}
        \caption{Vortex position variation.}\label{fig:tau_vs_position}
    \end{subfigure}
\caption{Effects of vortex generation parameters on vortex strength and vertical position. Shown are the variation of vortex circulation with pitch amplitude $\Delta\alpha_{\mathrm{eff}}$ and the variation of vortex vertical position with the shedding instance parametrized by $\tau_{\mathrm{v}}$.}
\label{fig:parameter_general_results}
\end{figure}

The influence of the flapping profile parameters on the vortex is summarized in Figures~\ref{fig:exp_vortex_strength}--\ref{fig:parameter_general_results}. Figures~\ref{fig:exp_vortex_strength} and \ref{fig:exp_vortex_position} present representative vorticity snapshots illustrating variations in vortex strength and transverse position associated with $\Delta\alpha_\mathrm{eff}$ and $\tau_\mathrm{v}$, respectively. Figure~\ref{fig:parameter_general_results} provides a quantitative overview of the vortex characteristics, including simulations and experimental cases at three Reynolds numbers. Systematic relationships between the flapping parameters and the resulting vortex characteristics are identified from these results, with consistent trends observed across numerical simulations and experiments.

\begin{enumerate}
    \item Vortex rotation orientation \\
    The rotation orientation of the generated vortex is directly controlled by the sign of the rapid pitch maneuver. Pitch-down motions (negative $\Delta \theta$ and $\Delta \alpha_{\mathrm{eff}}$) produce CW vortices, whereas pitch-up motions (positive $\Delta \theta$ and $\Delta \alpha_{\mathrm{eff}}$) produce CCW vortices. In addition, CCW vortices are accompanied by a trailing wake that extends ahead of the vortex core and impinges on the downstream airfoil prior to the primary vortex interaction. Conversely, the wake associated with CW vortices is oriented such that it does not impinge on the downstream airfoil prior to the arrival of the vortex.

    \item Vortex strength \\
    The strength of the generated vortex, characterized by its circulation $\Gamma^*$, increases monotonically with the magnitudes of $\Delta \theta$ and $\Delta \alpha_{\mathrm{eff}}$. This trend is consistent with classical pitching-airfoil behavior, in which the circulation is directly related to the pitch amplitude. As observed in the vorticity fields, increasing the magnitude of $\Delta\alpha_{\mathrm{eff}}$ leads to a corresponding increase in the magnitude of the vortex circulation, with the sign of $\Gamma$ determined by the sign of the rapid pitch maneuver. This dependence is clearly observed in Figure~\ref{fig:alpha_vs_circulation}, which shows a positive correlation between $\Gamma$ and $\Delta\alpha_{\mathrm{eff}}$. The increase in circulation gradually tapers off as $\Delta\alpha_{\mathrm{eff}}$ reaches higher amplitudes across all Reynolds numbers considered, presumably due to the airfoil approaching a stalling regime.
    \par
    Our results also show that for small values of $\Delta\alpha_\mathrm{eff}$, secondary vortices may form in both experiments and simulations, whereas such structures are not observed for larger values of $\Delta\alpha_\mathrm{eff}$. This behavior is illustrated in the experimental results shown in Figures~\ref{fig:exp_vortex_strength}i and \ref{fig:exp_vortex_strength}j, where a secondary vortex of comparable magnitude appears near the primary vortex. One contributing factor is that, for a fixed $St_\mathrm{v}$, smaller values of $\Delta\alpha_\mathrm{eff}$ correspond to a shorter rapid-pitch duration $t_\mathrm{d2}$ (see Eqn.~\ref{eqn:st_td2}). When the rapid pitching motion ends, the shear layer shedding does not cease immediately, and a portion of the remaining vorticity can roll up into a secondary vortex. Although this effect might be partially mitigated by selecting a smaller $St_\mathrm{v}$ to increase $t_\mathrm{d2}$, the resulting primary vortex would then be more stretched and more difficult to roll up into a compact structure.
    
    \item Vortex position \\
    The transverse position of the vortex can be adjusted independently of its core structure by varying the timing parameter $\tau_{\mathrm{v}}$, or equivalently the rapid-pitch onset time $t_{\mathrm{s2}}$. Changes in $\tau_{\mathrm{v}}$ shift the vertical location at which the vortex forms and therefore its downstream trajectory. Smaller values of $\tau_{\mathrm{v}}$ correspond to earlier vortex shedding and lower formation positions, whereas larger values lead to vortices forming higher in the transverse direction. This trend is quantified in Fig.~\ref{fig:tau_vs_position}, which shows an approximately linear dependence of vortex position on $\tau_{\mathrm{v}}$.
\end{enumerate}
While the vortex core size is relatively insensitive to variations in the flapping parameters considered here and remains comparable between simulations and experiments, slight differences are observed in the vorticity distribution surrounding the core. Experiments show somewhat more compact vortices, and appear to be consistently smaller in radius than in simulations, most notably at low $\Delta\alpha_\mathrm{eff}$ (see Figure~\ref{fig:exp_vortex_strength}). Additionally, the vortex core size shows negligible variation across the three Reynolds numbers tested experimentally in this study.
The primary effect of increasing Reynolds number is instead observed in the vortex strength, as shown in Figure~\ref{fig:alpha_vs_circulation}.
These observations suggest that, for a given generating airfoil size, the characteristic vortex size remains approximately fixed, and that substantially larger or smaller vortices would require a change in chord length of the vortex generating airfoil. 

\subsection{Induced lift on downstream airfoil}\label{ssec:vortex_induced_lift}
\begin{figure}[!htbp]
     \centering
     \begin{subfigure}[b]{0.32\textwidth}
         \centering
         \includegraphics[width=\textwidth]{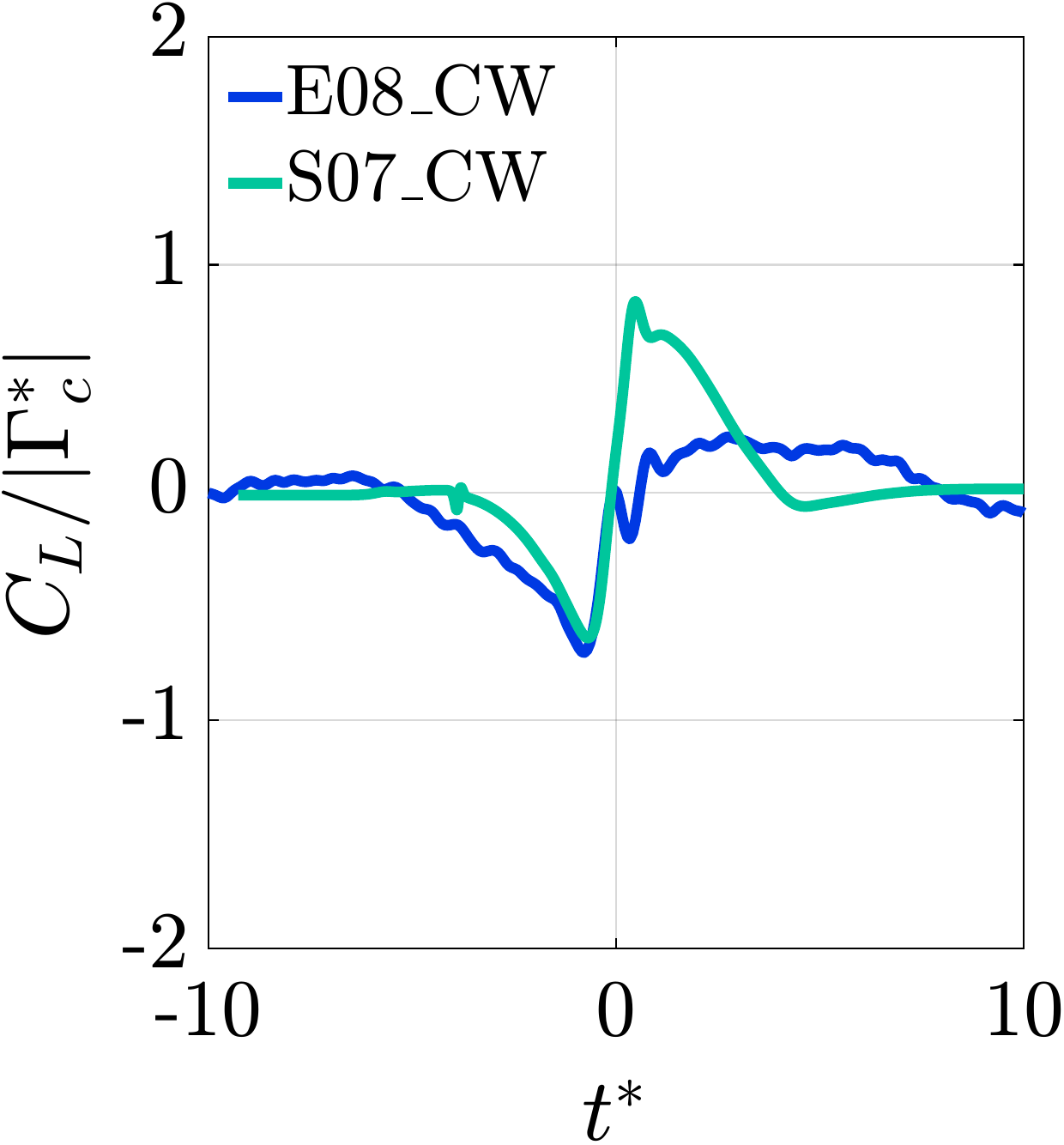}
         \caption{Below LE, $y_c^* \approx -0.5$}
         \label{fig:cl_low}
     \end{subfigure}
     \hfill
     \begin{subfigure}[b]{0.32\textwidth}
         \centering
         \includegraphics[width=\textwidth]{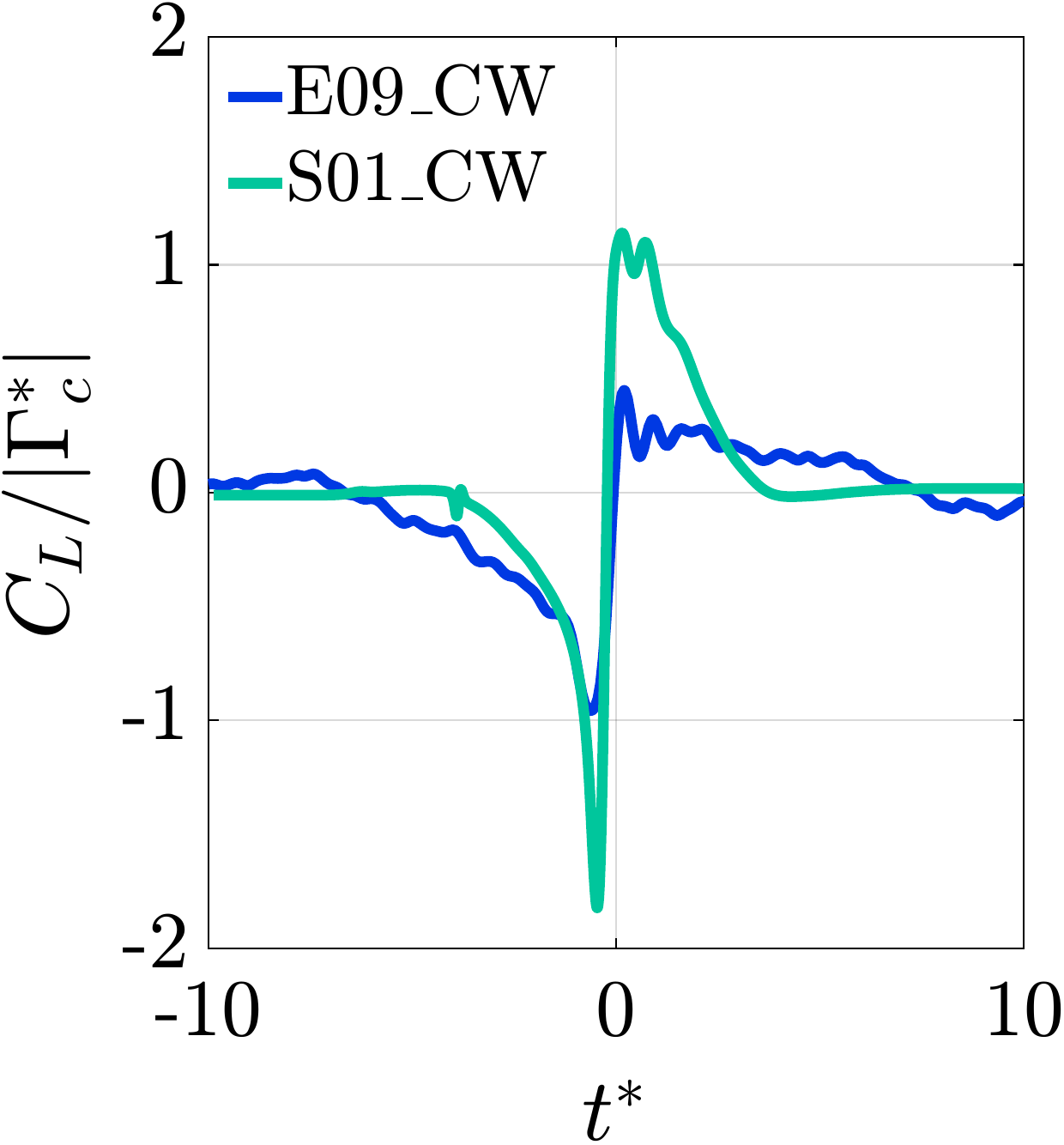}
         \caption{Near LE, $y_c^* \approx 0$}
         \label{fig:cl_mid}
     \end{subfigure}
     \hfill
     \begin{subfigure}[b]{0.32\textwidth}
         \centering
         \includegraphics[width=\textwidth]{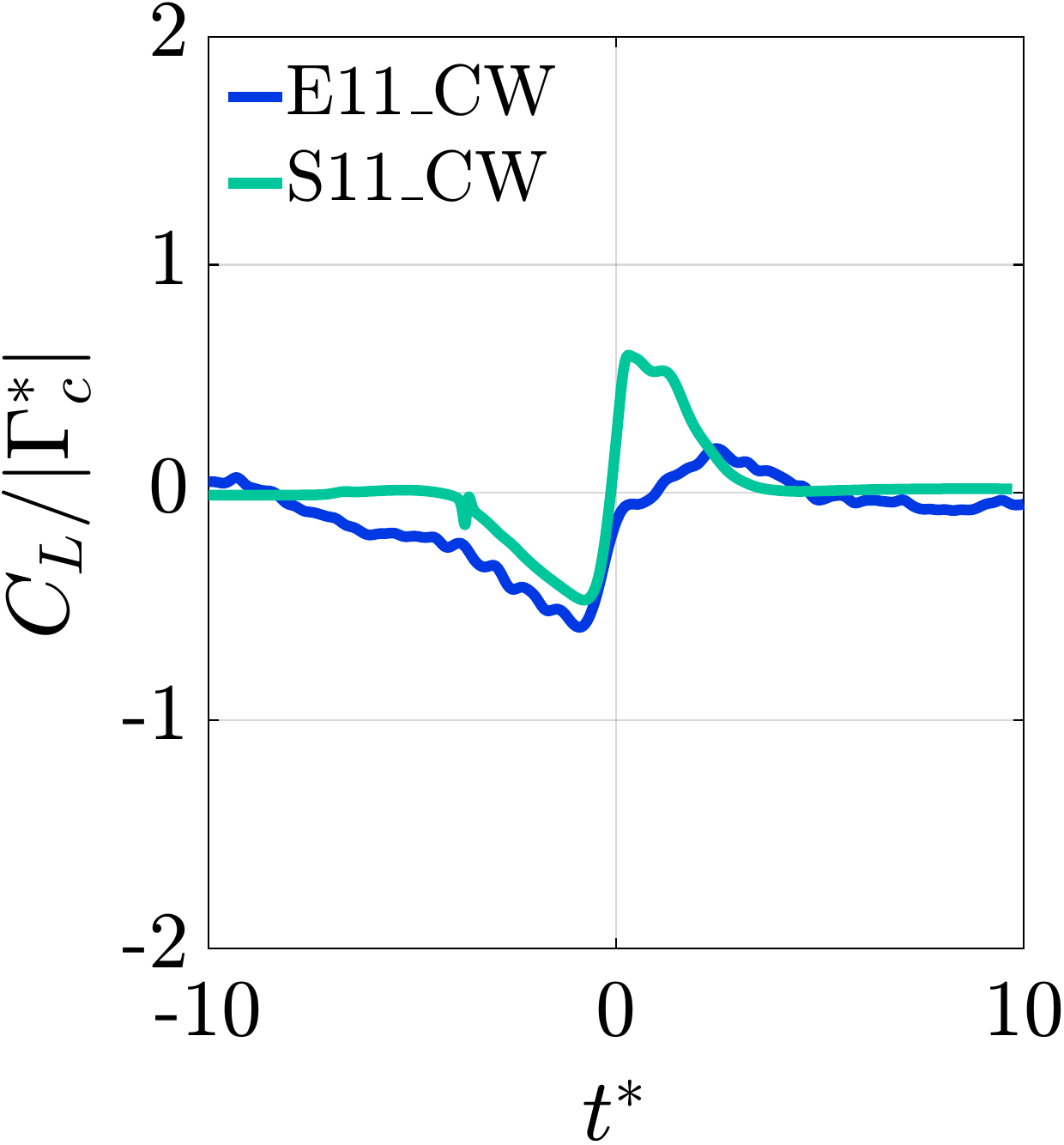}
         \caption{Above LE, $y_c^* \approx 0.5$}
         \label{fig:cl_high}
     \end{subfigure}
    \caption{Evolution of the lift coefficient ($C_L$) on the downstream airfoil during vortex encounters, normalized by the magnitude of vortex circulation. Results are shown for experiments and simulations for vortices located below, at, and above the leading edge (LE) of the downstream airfoil. Panel (b) corresponds to a direct impingement case. The non-dimensional time is defined such that $t^*=0$ corresponds to the instant when the vortex reaches the downstream LE.}
        \label{fig:cl}
\end{figure}

Although the primary focus of this study is the vortex generated by the flapping upstream airfoil, the characteristics of the vortex and how well it is isolated from the wake can be further assessed by examining the interaction with a downstream airfoil. Accordingly, the lift response of the downstream airfoil is examined for selected cases in which the vortex closely interacts with the airfoil. Figure~\ref{fig:cl} shows the time history of the lift coefficient ($C_L$) for representative CW vortices. The lift force is nondimensionalized using the chord length of the downstream airfoil, and the results are normalized by the vortex circulation to enable comparison between simulations and experiments. 
\par
As the clockwise vortex approaches the downstream airfoil, it induces a downwash near the leading edge, resulting in a gradual decrease in the lift coefficient to a negative peak. As the vortex convects over the airfoil, the lift rapidly increases to a positive maximum before gradually returning toward zero. This transient response is characteristic of vortex–airfoil interactions and is consistent with observations reported in previous studies \cite{barnes2018clockwise,hufstedler2019vortical,qian2023lift}. Among the configurations shown, the direct impingement cases produce the largest variation in $C_L$, as expected, whereas vortices passing above and below the leading edge yield comparable responses with smaller amplitudes.
\par
Relative to the experiments, the vortex-induced lift response in the simulations appears more temporally compact and exhibits a nearly symmetric fluctuation, consistent with trends reported by Zhong et al \cite{zhong2023sparse}. In the direct impingement case (Fig.~\ref{fig:cl_mid}), the overall variation in $C_L$ is significantly larger in the simulations, with both positive and negative peaks reaching higher amplitudes. In contrast, the opposite peak is noticeably flatter in all experimental cases, indicating that the vortex influence is more localized in time in the numerical results.
\par
These differences likely arise from the variation in Reynolds number that would result in faster vortex breakdown due to increased turbulence in experiments.
Additional discrepancies may also stem from differences in vortex-generator kinematics. In the simulations, the larger nondimensional heaving velocity produces a more obliquely oriented wake, shown in figure~\ref{fig:vortex_formation_comparison}, which reduces the duration over which wake-related effects influence the downstream airfoil.
Previous experimental studies, including Biler et al.~\cite{biler2019experimental} and Qian et al.~\cite{qian2022interaction}, have similarly reported asymmetric lift responses, in which the initial lift peak is dominant and subsequent fluctuations diminish as the vortex crosses the airfoil. 
\par
Importantly, the lift coefficient is near zero prior to the vortex encounter and gradually returns toward zero after the interaction, indicating that the downstream airfoil does not experience sustained loading from the extended wake. This behavior suggests that the long-term influence of the upstream wake is limited for the present flapping-foil configuration, in contrast to vortex-generation approaches based solely on pitching motions, where more persistent wake effects have been reported \cite{hufstedler2019vortical}.

\subsection{Implications for vortex gust design}\label{ssec:vortex_design}
Building on this parametric analysis, the relationships between the flapping profile parameters and the resulting vortex characteristics provide a framework for the systematic design of vortex gusts with prescribed rotation orientation, strength, and transverse position. Appropriate selection of the rapid pitch direction, pitch amplitude, and formation timing enables vortices with targeted properties to be generated in a controlled and physically interpretable manner.
\par
Based on these observations, a practical design strategy can be outlined. The desired vortex orientation and strength primarily determine the sign and magnitude of the rapid pitch parameters $\Delta \alpha_{\mathrm{eff}}$ and $\Delta \theta$, while the duration of the rapid pitching phase is selected within an appropriate Strouhal-number range to ensure compact vortex formation. Because this Strouhal number depends on the Reynolds number, its optimal range varies with the flow regime. The remaining pitch amplitudes and timing parameters, except for the vortex formation time $t_{\mathrm{s2}}$, exert comparatively weaker influence on the resulting vortex characteristics and therefore provide flexibility for accommodating experimental or numerical constraints. Care must nevertheless be taken to avoid excessively large effective angles of attack, which may induce undesired leading-edge separation. In addition, the vortex-generation process does not require the airfoil to initiate motion from a lower transverse position; reversing the direction of the heaving motion is equally feasible and may be advantageous when the overall structure of the vortex gust, including the extended wake, is of interest.
\par
Once the parameters governing vortex orientation and strength have been established, the transverse placement of the vortex can be adjusted through the formation timing, either via the normalized parameter $\tau_{\mathrm{v}}$ or equivalently the time instance $t_{\mathrm{s2}}$. As demonstrated in the preceding results, variations in $\tau_{\mathrm{v}}$ shift the vortex trajectory without significantly altering its internal structure, making this parameter a convenient final design choice.
\par
Additional manipulation of the vortex-generation process may be achieved by prescribing unequal values of $\Delta \alpha_{\mathrm{eff}}$ and $\Delta \theta$, or by introducing a phase offset between the rapid variations in pitch angle and effective angle of attack. Such adjustments become particularly relevant when $\Delta \alpha_{\mathrm{eff}}$ is small, as secondary vortices may become more pronounced under these conditions, as illustrated in Figures~\ref{fig:exp_vortex_strength}i and \ref{fig:exp_vortex_strength}j. Introducing a phase offset can partially mitigate the strength of these secondary vortices, while prescribing unequal amplitudes primarily modifies the strength of the primary vortex. Although these configurations can alter the overall vortex structure, systematically quantifying their effects would introduce additional degrees of freedom beyond the scope of the present study, since the phase offset that most effectively suppresses secondary vortices may vary with other profile parameters. Selected examples illustrating these effects are therefore presented in Appendix~\ref{app2}.

\section{Conclusion}\label{sec:conclusion}
In this study, a method for generating isolated vortex gusts with limited wake influence during horizontal impingement was introduced. The generated vortices propagate along trajectories that remain nearly parallel to the incoming flow, enabling well-controlled vortex–airfoil interactions. The rotation orientation, strength, and vertical position of the vortex were shown to vary systematically with selected parameters of the flapping profile. In addition, the associated wake extends obliquely from the primary vortex, thereby limiting its influence on the downstream airfoil, as evidenced by the measured lift response.
\par
Through appropriate selection of the profile parameters, vortex gusts with prescribed orientation, strength, and position can be generated within a well-defined and controllable range. Comparisons between numerical simulations and experiments demonstrate qualitatively consistent vortex formation behavior, although the rapid pitching kinematics required for vortex formation, characterized by the associated Strouhal number, depend on the Reynolds number. Despite this dependence, the overall trends in vortex characteristics are preserved across the Reynolds-number range examined.
\par
Future work will utilize the present method to investigate aeroelastic interactions between vortex gusts and passively moving downstream airfoils. In particular, we will examine the coupled dynamics of vortex-induced loading and structural response, with the goal of identifying strategies to alleviate aerodynamic loading and aeroelastic response through the design of structural parameters.

\section*{Acknowledgments}
This work is funded by AFOSR Award Number FA9550-23-1-0478 monitored by Dr. Gregg Abate. Computational time is provided by the Center for High-Throughput Computing (CHTC) at UW-Madison. AI has been used to polish the grammar and wording in this manuscript, but has not been used for generative text or figure modification. Conceptualization: JAF, KB; Methodology, Data Curation and Analysis: BY, EHC; Writing – original draft: BY, EHC; Writing – review \& editing: JAF, KB. 

\clearpage
\appendix

\section{Table of parameters for the vortex generation kinematic profile}
\label{app1}
\begin{table}[!h]
\centering
\caption{Kinematic parameters (Eqns. \ref{eqn:vortex_profile_theta} and \ref{eqn:vortex_profile_alpha}) for all cases presented in this paper. Case labels starting with ``S'' refer to simulations, while those starting with ``E'' refer to experiments. Simulations were performed at a $Re=1000$, and all experimental cases were done at $Re=28000$, $35000$, and $42000$.}
\label{tab:detailed_param}

\setlength{\tabcolsep}{4pt}
\renewcommand{\arraystretch}{0.85}

\begin{tabular}{|l|cccccccc|cccc|}
\hline
Case
& $t_{\mathrm{s1}}$
& $t_{\mathrm{d1}}$
& $t_{\mathrm{s2}}$
& $t_{\mathrm{d2}}$
& $t_{\mathrm{s3}}$
& $t_{\mathrm{d3}}$
& $St_{\mathrm{v}}$
& $\tau_{\mathrm{v}}$
& $\theta_0$
& $\alpha_{\mathrm{eff},0}$
& $\Delta\theta$
& $\Delta\alpha_{\mathrm{eff}}$\\
\hline
S01\_CW  & 3.00 & 1.50 & 5.83 & 0.35 & 6.85 & 1.50 & 1.634 & 0.665 & 40 & 6  & -15   & -15    \\
S02\_CW  & 3.00 & 1.50 & 5.83 & 0.24 & 6.81 & 1.50 & 1.589 & 0.643 & 40 & 6  & -10   & -10    \\
S03\_CW  & 3.00 & 1.50 & 5.81 & 0.16 & 6.75 & 1.50 & 1.787 & 0.627 & 40 & 6  & -7.5  & -7.5   \\
S04\_CCW & 3.00 & 1.50 & 6.64 & 0.15 & 7.34 & 1.50 & 1.907 & 0.796 & 25 & -7 & 7.5   & 7.5    \\
S05\_CCW & 3.00 & 1.50 & 6.75 & 0.20 & 7.34 & 1.50 & 1.907 & 0.852 & 25 & -7 & 10    & 10     \\
S06\_CCW & 3.00 & 1.50 & 6.72 & 0.32 & 7.34 & 1.50 & 1.787 & 0.881 & 25 & -7 & 15    & 15     \\
S07\_CW  & 3.00 & 1.50 & 5.23 & 0.35 & 6.85 & 1.50 & 1.634 & 0.365 & 40 & 6  & -15   & -15    \\
S08\_CW  & 3.00 & 1.50 & 5.53 & 0.35 & 6.85 & 1.50 & 1.634 & 0.515 & 40 & 6  & -15   & -15    \\
S10\_CW  & 3.00 & 1.50 & 6.13 & 0.35 & 6.85 & 1.50 & 1.634 & 0.815 & 40 & 6  & -15   & -15    \\
S11\_CW  & 3.00 & 1.50 & 6.43 & 0.35 & 6.85 & 1.50 & 1.634 & 0.965 & 40 & 6  & -15   & -15    \\
\hline
E01\_CW   & 6.00 & 3.818 & 12.473 & 0.873 & 16.00 & 3.82 & 0.655 & 0.5 & 26.5 &   9 & -15 & -15 \\
E02\_CW   & 6.00 & 3.818 & 12.618 & 0.582 & 16.00 & 3.82 & 0.655 & 0.5 & 23.5 &   6 & -10 & -10 \\
E03\_CW   & 6.00 & 3.818 & 12.763 & 0.291 & 16.00 & 3.82 & 0.655 & 0.5 & 20.5 &   3 &  -5 &  -5 \\
E04\_CCW  & 6.00 & 3.818 & 12.763 & 0.291 & 16.00 & 3.82 & 0.655 & 0.5 & 14.5 &  -3 &   5 &   5 \\
E05\_CCW  & 6.00 & 3.818 & 12.618 & 0.582 & 16.00 & 3.82 & 0.655 & 0.5 & 11.5 &  -6 &  10 &  10 \\
E06\_CCW  & 6.00 & 3.818 & 12.473 & 0.873 & 16.00 & 3.82 & 0.655 & 0.5 &  8.5 &  -9 &  15 &  15 \\
E07\_CW  & 6.00 & 3.818 & 10.880 & 0.873 & 16.00 & 3.82 & 0.655 & 0.2 & 26.5 &   9 & -15 & -15 \\
E08\_CW  & 6.00 & 3.818 & 11.942 & 0.873 & 16.00 & 3.82 & 0.655 & 0.4 & 26.5 &   9 & -15 & -15 \\
E09\_CW  & 6.00 & 3.818 & 13.004 & 0.873 & 16.00 & 3.82 & 0.655 & 0.6 & 26.5 &   9 & -15 & -15 \\
E10\_CW  & 6.00 & 3.818 & 14.065 & 0.873 & 16.00 & 3.82 & 0.655 & 0.8 & 26.5 &   9 & -15 & -15 \\
E11\_CW  & 6.00 & 3.818 & 15.127 & 0.873 & 16.00 & 3.82 & 0.655 & 1.0 & 26.5 &   9 & -15 & -15 \\
E12\_CW  & 6.00 & 3.818 & 15.127 & 0.873 & 16.00 & 3.82 & 0.655 & 0.0 & 26.5 &   9 & -15 & -15 \\
E13\_CCW & 6.00 & 3.818 &  9.818 & 0.873 & 16.00 & 3.82 & 0.655 & 0.0 &  8.5 &  -9 &  15 &  15 \\
E14\_CCW & 6.00 & 3.818 & 15.127 & 0.873 & 16.00 & 3.82 & 0.655 & 1.0 &  8.5 &  -9 &  15 &  15 \\
E15\_CW  & 6.00 & 3.818 & 12.327 & 1.164 & 16.00 & 3.82 & 0.655 & 0.5 & 29.5 &  12 & -20 & -20 \\
E16\_CCW & 6.00 & 3.818 & 12.327 & 1.164 & 16.00 & 3.82 & 0.655 & 0.5 &  5.5 & -12 &  20 &  20 \\
\hline
\end{tabular}
\end{table}

\section{Extra manipulation of the kinematic profile}
\label{app2}

A more general form of the vortex-generating profile includes a parameter $t_{\mathrm{off}}$, which denotes the time offset between the onset of the rapid pitch maneuver and the change in the effective angle of attack. It results in a modified the $\alpha_{\mathrm{eff}}$ equation, 
\begin{equation}
\alpha_{\mathrm{eff}}(t) =
\alpha_{\mathrm{eff},0} \, r(t; t_{\mathrm{s1}}, t_{\mathrm{d1}})
+ \Delta \alpha_{\mathrm{eff}} \, r(t; t_{\mathrm{s2}}+t_{\mathrm{off}}, t_{\mathrm{d2}})
- (\alpha_{\mathrm{eff},0} + \Delta \alpha_{\mathrm{eff}}) \, r(t; t_{\mathrm{s3}}, t_{\mathrm{d3}}).
\end{equation}

The remaining parameters follow the conventions introduced in \S~\ref{sec:generation}. 

\begin{figure}[!htbp]
     \centering
     \begin{subfigure}[b]{\textwidth}
         \centering
         \includegraphics[width=0.28\textwidth]{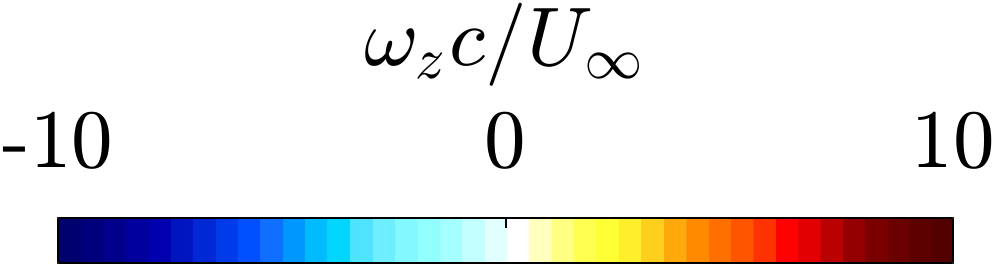}
     \end{subfigure}
     \hfill
     \begin{subfigure}[b]{0.32\textwidth}
         \centering
         \includegraphics[width=0.97\textwidth]{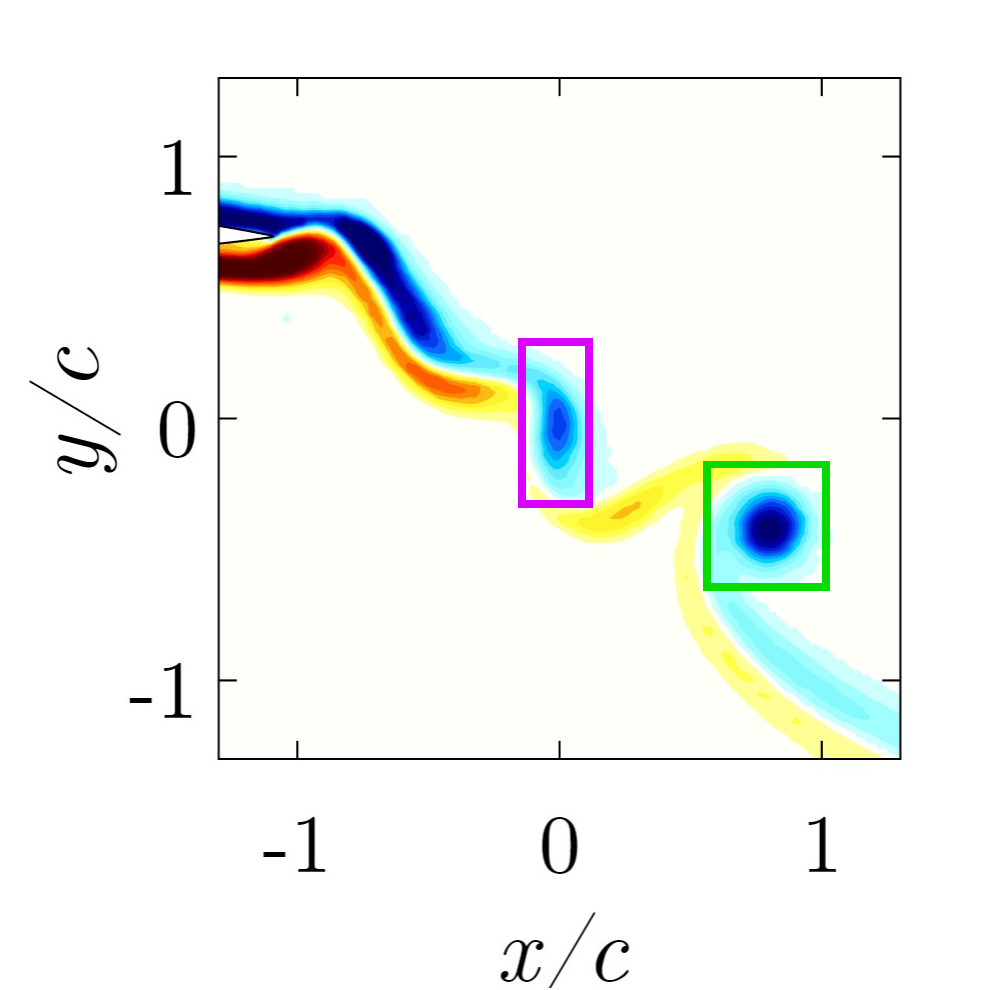}
         \caption{$t_{\mathrm{off}}^*=0,\;\Gamma_{\mathrm{sec}}^*\approx -0.21$}
     \end{subfigure}
     \hfill
     \begin{subfigure}[b]{0.32\textwidth}
         \centering
         \includegraphics[width=0.97\textwidth]{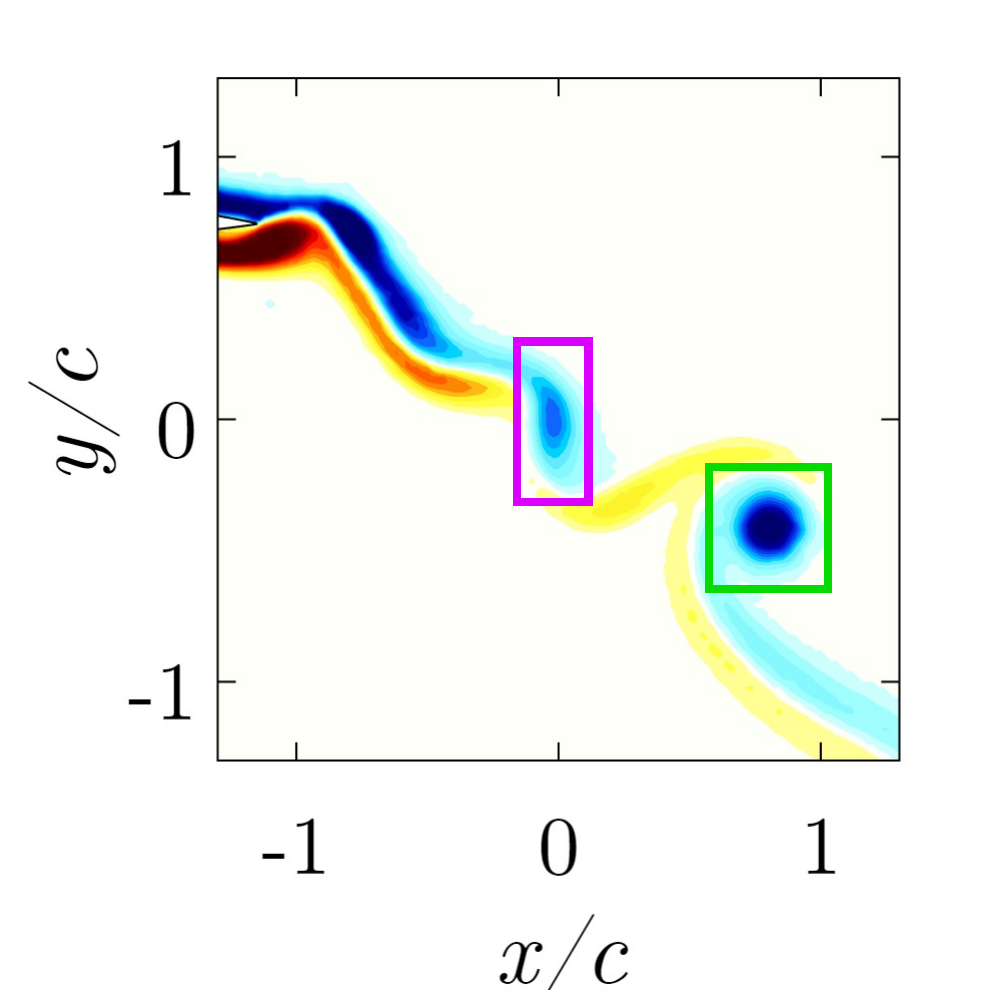}
         \caption{$t_{\mathrm{off}}^*=-0.1,\;\Gamma_{\mathrm{sec}}^*\approx -0.18$}
     \end{subfigure}
     \hfill
     \begin{subfigure}[b]{0.32\textwidth}
         \centering
         \includegraphics[width=0.97\textwidth]{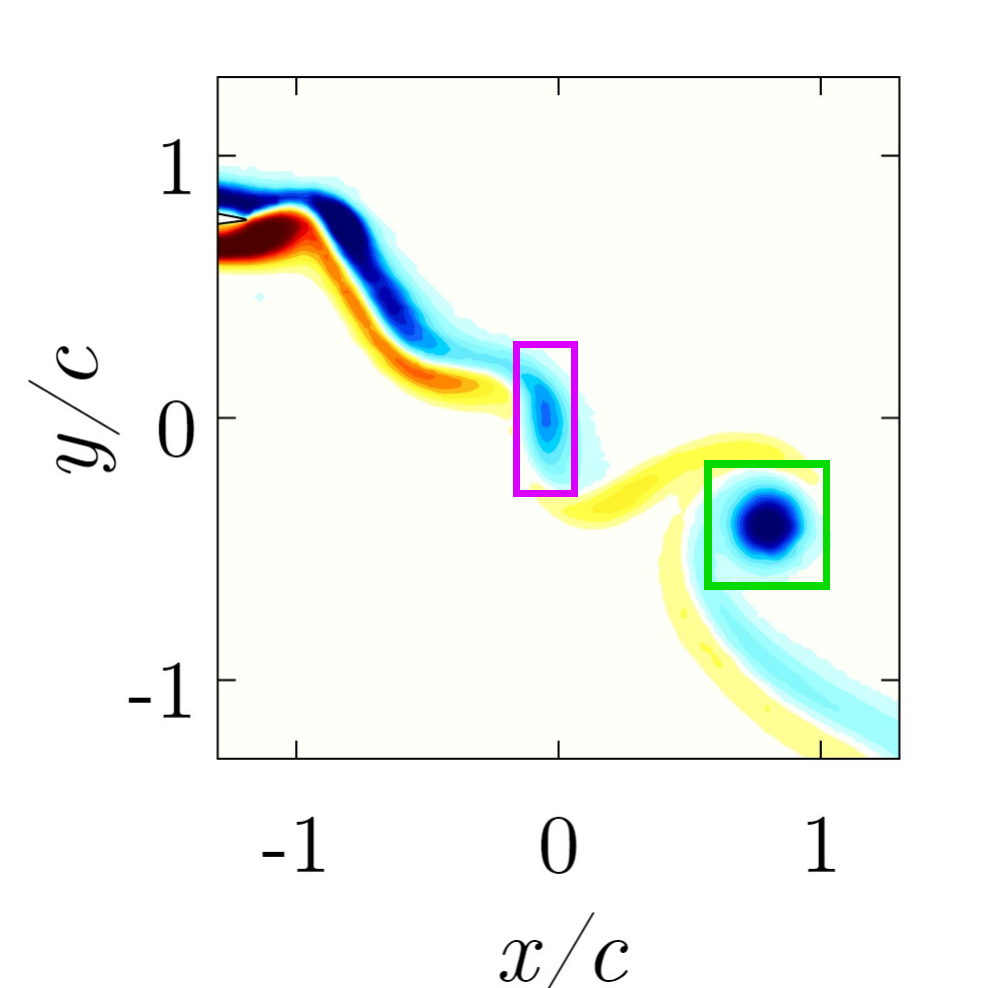}
         \caption{$t_{\mathrm{off}}^*=-0.2,\;\Gamma_{\mathrm{sec}}^*\approx -0.17$}
     \end{subfigure}
    \caption{Vorticity contours illustrating the effect of the time offset $t_{\mathrm{off}}$ between the rapid variations of $\theta$ and $\alpha_{\mathrm{eff}}$. Primary vortices are highlighted by green boxes, and secondary vortices by magenta boxes. The non-dimensional circulation of the secondary vortex is denoted by $\Gamma_{\mathrm{sec}}^*$. The origin has been shifted to be close to the center of the secondary vortex.}
        \label{fig:offset}
\end{figure}

As discussed in Section~\ref{sec:result}, small values of $\Delta\theta$ and $\Delta\alpha_{\mathrm{eff}}$ may lead to the formation of secondary vortices alongside the primary vortex. A representative example is case $S03\_CW$, which was also shown in Figure~\ref{fig:exp_vortex_strength}. In that figure the visualization was scaled to emphasize the primary vortex, whereas Figure~\ref{fig:offset}(a) shows the same case with a wider view that reveals the accompanying secondary vortex. Figures~\ref{fig:offset}(b)--(c) illustrate the effect of introducing a negative time offset $t_{\mathrm{off}}$, which advances the rapid variation of $\alpha_{\mathrm{eff}}$ relative to the pitch motion. All three cases shown here differ only in the value of $t_{\mathrm{off}}$, with all other parameters kept identical. With increasing magnitude of the offset, the secondary vortex becomes weaker, as demonstrated by the decreasing magnitude of its circulation from 0.21 to 0.17 (a 19\% decrease in strength). However, this adjustment only partially alleviates its formation and does not eliminate it entirely. This approach may nevertheless be useful in situations where small values of $\Delta\theta$ and $\Delta\alpha_{\mathrm{eff}}$ are required and the presence of a weak secondary vortex is acceptable.

\bibliography{reference}
\end{document}